\documentclass[aps,nofootinbib,notitlepage,longbibliography,twocolumn, superscriptaddress]{revtex4-1}
\usepackage{amsmath,amssymb}
\usepackage{slashed}
\usepackage{graphicx}
\usepackage{bm}
\usepackage{float}
\usepackage[T1]{fontenc}
\usepackage[utf8]{inputenc}
\usepackage{gauss} 
\usepackage[top=1.0in,bottom=1.0in,left=1.0in,right=1.0in]{geometry}
\usepackage[colorlinks,linkcolor=blue,citecolor=blue]{hyperref}
\usepackage{subfig}
\newcommand{\be}{\begin{equation}}
\newcommand{\ee}{\end{equation}}
\newcommand{\bea}{\begin{eqnarray}}
\newcommand{\eea}{\end{eqnarray}}
\newcommand{\ba}{\begin{eqnarray}}
\newcommand{\ea}{\end{eqnarray}}

\allowdisplaybreaks

\def\be{\begin{eqnarray}}
\def\ee{\end{eqnarray}}
\def\bea{\be}
\def\eea{\ee}

\def\roughly#1{\mathrel{\raise.3ex\hbox{$#1$\kern-.75em%
\lower1ex\hbox{$\sim$}}}}

\usepackage{lipsum}

\date{\today}
\begin{abstract}
Following on our recent analysis of the energy momentum tensor (EMT) of light nuclei in the impulse approximation, we evaluate the leading exchange corrections also upto momenta of the order of the nucleon mass. The exchange contributions to the EMT, are composed of the pair interaction, plus the seagull and the pion exchange interactions, modulo the recoil correction. The exchange
contributions are shown to satisfy the current conservation requirement. These contributions are  small  compared to those from the impulse approximation for most of the deuteron gravitational 
form factors (GFFs), for momenta smaller than half of the nucleon mass. For larger momenta, the exchange contributions are significant for the deuteron A- and D-GFFs. We suggest that the pion GFFs can be extracted from the exchange contributions of select deuteron GFFs.
\end{abstract}
\begin{document}
\title{Deuteron gravitational form factors: \\
exchange currents
}
\author{Fangcheng He}
\email{fangcheng.he@stonybrook.edu}
\affiliation{Center for Nuclear Theory, Department of Physics and Astronomy,
Stony Brook University, Stony Brook, New York 11794–3800, USA}

\author{Ismail Zahed}
\email{ismail.zahed@stonybrook.edu}
\affiliation{Center for Nuclear Theory, Department of Physics and Astronomy,
Stony Brook University, Stony Brook, New York 11794–3800, USA}
\maketitle

\section{Introduction}
The lightest and simplest nuclear system is the deuteron, with binding energy
(2.225 MeV), charge radius (2.16 fm)  and magnetic moment (0.857 in Bohr magnetons). The deuteron large size and weak binding, suggests that its shape is largely determined by the exchange of a pion.

The gravitational form factors (GFFs) of the deuteron carry important information on its mass and spin distribution. The deuteron GFFs provide important  insights to the fundamental gravitational content of a light nucleus. Model calculations of the deuteron GFFs include the lightcone convolution model~\cite{Freese:2022yur} and more recently the Skyrme model~\cite{GarciaMartin-Caro:2023klo}, et al. 
Empirically, the gluon GFFs of light nuclei maybe accessible  through  threshold electromagnetic production of heavy mesons (charmonium, bottomonium)  off nuclei. 

The nucleus is composed of nucleons (protons and neutrons) bound by strong QCD interactions. Most of what is known about nuclei, follow from  their electromagnetic properties using elastic electron scattering at intermediate energies, where the nucleons appear as rigid but extended bodies exchanging mesons, albeit mostly pions. The disparity between the fundamental and unconfined degrees of QCD (quarks and gluons) and the observed but confined 
degrees of freedom (mesons and nucleons) necessitates the exploration of innovative probing methods. The ultimate goal is to unravel the composition of the nucleons, and the interactions between nucleons in a nucleus.

Recently, we investigated the deuteron GFFs using the impulse approximation,  with momentum transfer  in the nucleon mass range, i.e. $k\sim m_N$~\cite{He:2023ogg}. We found large deviations of the deuteron  GFFs with that of the nucleon. One the other hand, it is well established that the meson exchange-current contributions to the electromagnetic form factors of two- and three-nucleon systems are not negligible~\cite{Hockert:1973fot,Chemtob:1974nf,Kloet:1973mj,Jackson:1975fys} at large momentum transfer, i.e. 
$k\geq \frac 12 m_N$.
The motivation of this work is to analyse the corrections induced by  pion exchange currents to the deuteron GFFs.

This paper is a follow up on our recent analysis of the EMT of light nuclei in the impulse approximation~\cite{He:2023ogg}. In section~\ref{SEC_EMTX} we discuss
the Born exchange corrections to the impulse approximation, where the relativistic contributions are first retained. To leading order in the pion-nucleon
pseudo-vector coupling, the exchange contributions are shown to satisfy the EMT current conservation condition on general ground. This is explicitly checked in the non-relativistic limit. In section~\ref{SEC_III} we detail the exchange contributions stemming from the non-relativistic reduction of the nucleon pair
exchange, the seagull and pion exchanges, as well the recoil correction. In section~\ref{SEC_IV} we compare the exchange contributions to the impulse approximation for the deuteron EMT. Our conclusions are in section~\ref{SEC_CON}.

\section{Deuteron exchange EMT}
\label{SEC_EMTX}
The corrections to the impulse approximation will be sought in the Born corrections at tree level. 
The results will be checked to satisfy the EMT conservation law. The same observation holds for 
the non-relativistic limit of the extracted exchange currents.

\begin{figure*}
\begin{center}
\includegraphics[scale=0.6]{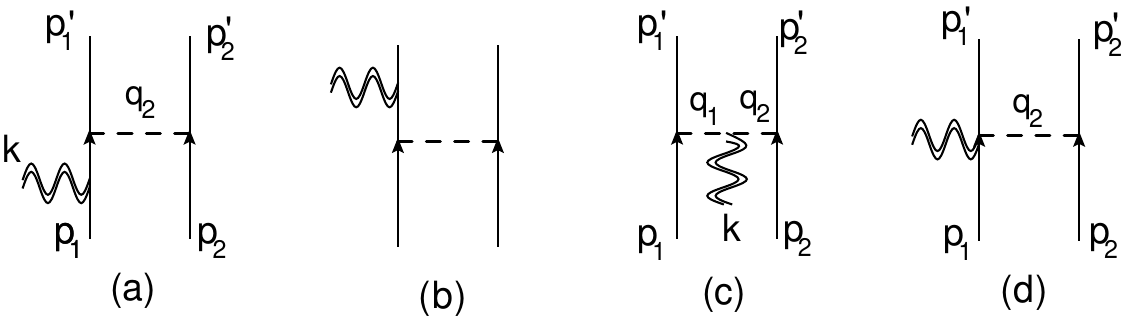}
\caption{
Exchange contributions to the EMT: (a+b) pair contribution; (c) pion exchange contribution; (d) Seagull contribution. The solid, dashed and wavy lines represent the nucleon, pion and graviton, respectively.} 
\label{fig:XPI}
\end{center}
\end{figure*}

\subsection{Born approximation}
\label{sec:Born_app}
The exchange contributions to the EMT in the Born approximation, are shown in Fig.~\ref{fig:XPI}. Using the canonical pseudo-vector coupling for the pion, the contributions $(a+b)$ in Fig.~\ref{fig:XPI} amount to
\begin{widetext}
\be
T^{\mu\nu}_{X\pi}(k, 1,2)=&&-
\bigg(\frac{g_{\pi N}}{2m_N}\bigg)^2
\bar{u}(p_1')\bigg(
\slashed{q}_2\gamma_5\tau_1^a S_F(p_1+k)
\,T_N^{\mu\nu}(p_1+k,p_1)+
T_N^{\mu\nu}(p_1', p_1'-k)\,S_F(p_1^\prime-k)\slashed{q}_2\gamma_5\tau_1^a\bigg) u(p_1)
\,\nonumber\\
&&\qquad\qquad\,\,\,\times \Delta_\pi(q_2^2)
\,\bar{u}(p_2')\slashed{q}_2\gamma_5\tau_2^au(p_2) + (1\leftrightarrow 2)
\ee
where the coupling constant $g_{\pi N}$ is  related to the nucleon axial charge $g_A$ through the Goldberger-Treiman relation  $g_{\pi N}/m_N=g_A/f_\pi$, with the pion decay constant $f_\pi=93$ MeV. Here $S_F$ and $\Delta_\pi$ are  the nucleon and pion propogators with $S_F(l)=i/(\slashed{l}-m_N+i0)$ and $\Delta_\pi(q_{1,2})=i/(q_{1,2}^2-m_\pi^2+i0)$.
The nucleon EMT $T_N^{\mu\nu}(l_2,l_1)$ is
\be
\label{A1tmp}
T_N^{\mu\nu}(l_2,l_1)=
A(k)\gamma^{(\mu}P_l^{\nu)}+B(k)\frac{iP_l^{(\mu}\sigma^{\nu)\alpha}k_\alpha}{2m_N}+C(k)\frac{k^\mu k^\nu-\eta^{\mu\nu}k^2}{m_N}-g^{\mu\nu}\left(\frac{\slashed{l}_1+\slashed{l}_2}{2}-m_N\right)
\ee
where $P_l=\frac{l_1+l_2}{2}$. The last term is the off-shell contribution.  In  leading order in 
chiral perturbation theory (ChPT), $A(k)=1$, $B(k)=C(k)=0$.~\cite{Alharazin:2020yjv}. Away from the chiral point, the nucleon GFFs have been extracted
from experiment~\cite{Duran:2022xag}, with overall agreement with QCD lattice simulations~\cite{Hackett:2023rif}, pQCD factorisation~\cite{Guo:2023pqw} 
and dual gravity~\cite{Mamo:2019mka,Mamo:2022eui}.
The contribution (c) with the intermediate pion EMT contributes
\be
T^{\mu\nu}_{X\pi\pi}(k, 1, 2)=-\bigg(\frac{g_{\pi N}}{2m_N}\bigg)^2T_{\pi}^{\mu\nu}( q_2,q_1)\bar{u}(p_1')\slashed{q}_1\gamma_5\tau_1^a u(p_1)\Delta_\pi(q_1)\Delta_\pi(q_2)
\bar{u}(p_2')\slashed{q}_2\gamma_5\tau_2^a u(p_2)
\ee

The general pion EMT can be written as
\be\label{eq:onshellpi}
T_\pi^{\mu\nu}(q_2,q_1)=\langle \pi, q_2|T_\pi^{\mu\nu}(0)|\pi, q_1\rangle=
\frac 12 (g^{\mu\nu}k^2-k^\mu k^\nu)\,
T_{1\pi}(k)+ \frac 12 l^\mu l^\nu\,
T_{2\pi}(k)-\frac 12 g^{\mu\nu}(q_2^2+q_1^2-2m_\pi^2)
\ee
where $l^\mu=(q_1+q_2)^\mu$ and $k^\mu=(q_2-q_1)^\mu$,
with the pion invariant form factors $T_{1\pi}(k)=T_{2\pi}(k)=1$ in  leading order in ChPT~\cite{Alharazin:2020yjv} . The last seagull contribution (d) to the EMT is
\be\label{eq:TXS}
T_{XS}^{\mu\nu}(k, 1, 2)=&&i
\bigg(\frac{g_{\pi N}}{2m_N}\bigg)^2
\bigg(\frac 12 g^{\mu\lambda}q_{2}^\nu +
\frac 12 g^{\nu\lambda}q_{2}^\mu-g^{\mu\nu}q^\lambda_2\bigg)\nonumber\\
&&\qquad\qquad\times\bar{u}(p_1')\gamma_\lambda \gamma_5\tau_1^a u(p_1)
\,\Delta_\pi(q_2)
\bar{u}(p_2')\slashed{q}_2\gamma_5\tau_2^a u(p_2) +(1\leftrightarrow 2)
\ee
To the order considered, we have checked that the  $g^{\mu\nu}$ contributions in (\ref{A1tmp}) and (\ref{eq:TXS}) cancel out.
They will not be considered in what follows.

\subsection{EMT current conservation}
The conservation of the EMT exchange contributions in the Born approximation and to leading order, amount to
\be\label{eq:EMT_conse}
k_\mu\big(
T^{\mu\nu}_{X\pi}(k, 1,2)+
T^{\mu\nu}_{X\pi\pi}(k, 1,2)+
T^{\mu\nu}_{XS}(k, 1,2)
\big)={\cal O}\bigg(\frac{g_{\pi N}^3}{m^3_N}\bigg)
\ee
\end{widetext}
which is seen to follow readily for the on-shell leg of each of  the first two contributions, and after some algebra for the off-shell parts  plus the seagull contributions. Note that in Eq.~(\ref{eq:EMT_conse}), 
the nucleon and pion EMT are fixed by ChPT.

The contributions in Fig.~\ref{fig:XPI}  (a+b) contain the positive frequency plus the negative frequency nucleon contribution in the intermediate state say of the deuteron. The former is part of the deuteron potential and should be removed from the contribution to the exchange current operator. With this in mind, we split the Feynman intermediate propagator 
\be
S_F(P)=S_F^+(P)+S_F^-(P)
\ee
We now define the exchange EMT current by removing the nucleon state
\be
T^{\mu\nu}_X=T^{\mu\nu\,-}_{X\pi}(k, 1,2)+
T^{\mu\nu}_{X\pi\pi}(k, 1,2)+
T^{\mu\nu}_{XS}(k, 1,2)\nonumber\\
\ee
which is seen to satisfy the modified EMT conservation 
\begin{widetext}
\be
\label{DX20}
k_\mu T^{\mu\nu}_{X}\equiv 
k_\mu\big(T^{\mu\nu\,-}_{X\pi}(k, 1,2)+
T^{\mu\nu}_{X\pi\pi}(k, 1,2)+
T^{\mu\nu}_{XS}(k, 1,2)\big)=
-k_\mu\,T^{\mu\nu\,+}_{X\pi}(k, 1,2)
\ee
More specifically, for in-out on-shell nucleons, we have
\bea
k_\mu\,T^{\mu\nu\,+}_{X\pi}(k, 1,2)&=&
-T_N^{0\nu}(p'_1, p_1'-k)V_\pi(p_1'-k, p_1; p_2', p_2)+V_\pi(p_1', p_1+k;p_2', p_2)T_N^{0\nu}(p_1+k, p_1)
\nonumber\\
&+& 1\leftrightarrow 2
\eea
with the tree-level pseudo-vector pion exchange potential
\be
V_\pi(p_1', p_1; p_2', p_2)=-i\bigg(\frac{g_{\pi N}}{2m_N}\bigg)^{2}
\bar{u}(p_1')\slashed{q}_1\gamma_5\tau^au(p_1)\,\Delta_\pi(q_2)\,
\bar{u}(p_2')\slashed{q}_2\gamma_5\tau^a\,u(p_2) 
\ee
\end{widetext}
Hence, the right-hand-side of (\ref{DX20}) is emanable to the commutator as a convolution in momentum space,  of the pion exchange potential with the EMT current
\be
k_\mu\,T^{\mu\nu\,+}_{X\pi}
=[V_\pi, T_N^{0\nu}]
\ee
We now separate the  time component of the exchange current 
and define
the Hamiltonian $$H=T+V_\pi$$ as the sum of single nucleon kinetic contribution  $T$,  plus the pair nucleon  pion  exchange contribution $V_\pi$,
with  the continuity equation
\be
\label{NREMT}
k^iT_X^{i\nu}
-[T, T_X^{0\nu}]-[V_\pi, T_N^{0\nu}]\sim 0
\ee
In the non-relativistic limit, the second contribution
is subleading, and (\ref{NREMT}) is satisfied by the leading contributions
\be
\label{NREMTX}
k^iT_X^{i\nu}
-[V_\pi, T_N^{0\nu}]\sim 0
\ee

\begin{figure*}
\begin{center}
\includegraphics[scale=0.6]{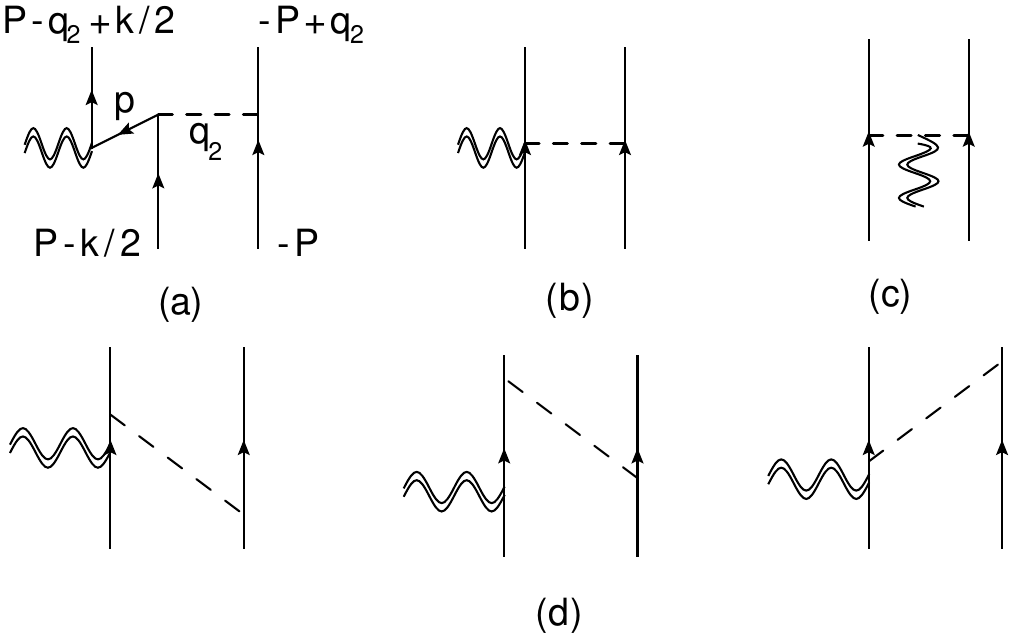}
\caption{
Exchange contributions to the EMT: (a)  The pair contribution, (b) seagull contribution and (c) the pion exchange contribution; (d) refers to  the recoil plus  wave function renormalization contributions. The solid, dashed and wavy lines represent the nucleon, pion and graviton, respectively.} 
\label{fig:NR_diagrams}
\end{center}
\end{figure*}

\section{Non-relativistic reduction}
\label{SEC_III}
To extract the non-relativistic contributions to the exchange currents, we will make use of the diagrams in~Fig.~\ref{fig:NR_diagrams}, with the positive nucleon contribution in (a)
subtracted, and the recoil contribution and wavefunction renormalization
in (d) retained. Again, the wiggly line refers to the insertion of the EMT,
and the dashed line to the pion exchange contribution
using the pseudo-vector coupling.

\subsection{Fig.~\ref{fig:NR_diagrams}a : Pair contribution}
The non relativistic reduction for  Fig.~(\ref{fig:NR_diagrams})a is 
\begin{widetext}
\bea
\label{TXPI}
T_{X\pi}^{00, ij}(k)&=&{\cal O}\bigg(\frac{g_{\pi N}^2}{m_N^4}\bigg)\nonumber\\ 
T_{X\pi}^{0j}(k)
&=&A(k)\tau_1\cdot\tau_2\,\frac{g_{\pi N}^2}{(2m_N)^2}\frac{\sigma_2.q_2 (-i (k\times q_2)^j-q_2^j \sigma_1.P_d+\sigma_1.q_2 P_d^j-\sigma_1^j P_d.q_2)}{4 m_Nw_{q_2}^2}
\nonumber\\
&+&(1\leftrightarrow 2)+\bigg(\frac{g_{\pi N}^2}{m_N^4}\bigg)
\eea
with $P_d=p_1+p_1'=2P-q_2$ and $w_{q_2}=(\vec{q}_2^{\,2}+m_\pi^2)^{1/2}$.
To evaluate the contributions (\ref{TXPI}) in a deuteron state, we follow our recent analysis in~\cite{He:2023ogg}. In particular, we have
\bea
\label{PAIR0J}
&&\left<+\frac{k}{2} m'\bigg|T_{X\pi}^{0j}\bigg|-\frac{k}{2} m\right>\nonumber\\
&=&\int d^3P\frac{d^3q_2}{(2\pi)^3}\Phi_{m'}^\dag\bigg(\vec{P}+\frac{\vec{k}}{4}-\vec{q_2}\bigg) T_{X\pi}^{0j}\Phi_m\bigg(\vec{P}-\frac{\vec{k}}{4}\bigg) \nonumber\\
&=&\int d^3P\frac{d^3q_2}{(2\pi)^3}d^3r_1d^3r_2e^{i(\vec{P}+\frac{\vec{k}}{4}-\vec{q_2}).\vec{r}_1}e^{-i(\vec{P}-\frac{\vec{k}}{4}).\vec{r}_2} \varphi^\dagger_{m'}(r_1)T_{X\pi}^{0j}\varphi_m(r_2)\nonumber\\
&=&\frac{\langle m'|(\vec{S}\times i\vec{k})^j|m\rangle}{2}J_{X\pi}
\eea
where the form factor $J_{X\pi}(k)$ can be expressed as
\bea\label{eq:FF_JXpi}
J_{X\pi}(k) &=&\frac{1}{8 \pi  m_N }A(k)\tau_1\cdot\tau_2\,\frac{g_{\pi N}^2}{(2m_N)^2}
\int \frac{dr}r\Bigg\{m_{\pi }^2 \bar{Y}_1 \left(2 u^2-\sqrt{2} u w-2 w^2\right) j_0\left(\frac{k r}{2}\right)
\nonumber\\
&-&\frac{j_1\left(\frac{k r}{2}\right)}{kr}m_{\pi }^2\bigg[4 m_{\pi } r u^2 Y_2-w \left(3 \sqrt{2} m_{\pi } r^2 u' Y_2+m_{\pi } r w Y_2-6 \sqrt{2} r u' \bar{Y}_1+15 w \bar{Y}_1\right)
\nonumber\\
&+&\sqrt{2} u \left(w \left(10 m_{\pi } r Y_2-24 \bar{Y}_1\right)+3 r w' \left(m_{\pi } r Y_2-2 \bar{Y}_1\right)\right)\bigg]\Bigg\},\nonumber\\
\eea
\end{widetext}
where $u$ and $w$ represent the reduced radial components 
${}^3S_1$ and ${}^3D_1$ of the deuteron wavefunction, which have been solved in our previous work~\cite{He:2023ogg},
\bea
Y_0&=&e^{-m_\pi r}/(m_\pi r)\nonumber\\
\bar{Y}_1&=&(1+1/(m_\pi r))Y_0\nonumber\\
Y_2&=&(3/(m_\pi r)^2+3/(m_\pi r)+1)Y_0
\eea
following the conventions in~\cite{Jackson:1975fys} for the 
electromagnetic current.

\subsection{Fig.~\ref{fig:NR_diagrams}b : Seagull contribution }
The non relativistic reduction for the seagull contribution in Fig.~\ref{fig:NR_diagrams}b,  
reads
\begin{widetext}
\bea
\label{eq:tmunuKR}
T_{XS}^{00}&=&{\cal O}\bigg(\frac{g_{\pi N}^2}{m_N^4}\bigg)
\nonumber\\
T_{XS}^{0j}&=&\tau_1\cdot\tau_2\frac{g_{\pi N}^2}{(2m_N)^2}\left(\frac{ \sigma_1^j \vec{P}_d.\vec{q}_2  \vec{\sigma}_2.\vec{q}_2}{4m_N w_{q_2}^2}-\frac{ q_2^j \sigma_1.P_d\sigma_2.q_2}{4m_N w_{q_2}^2}\right)
+(1\leftrightarrow 2)+{\cal O}\bigg(\frac{g_{\pi N}^2}{m_N^4}\bigg)
\nonumber\\
T_{XS}^{ij}&=&-\tau_1\cdot\tau_2\frac{g_{\pi N}^2}{(2m_N)^2}\frac{q_2^i \sigma_1^j+q_2^j \sigma_1^i}{2 w_{q_2}^2}\vec{\sigma}_2.\vec{q}_2+(1\leftrightarrow 2)+{\cal O}\bigg(\frac{g_{\pi N}^2}{m_N^4}\bigg)
\eea
The $t^{0j}$ contribution to the deuteron EMT, is
\bea
\left<+\frac{k}{2} m'\bigg|T_{XS}^{0j}\bigg|-\frac{k}{2} m\right>&=&\frac{\langle m'|(\vec{S}\times i\vec{k})^j|m\rangle}{2}J_{XS}
\eea
where the form factor $J_{XS}$ can be expressed as
\begin{subequations}
\bea
\label{eq:FF_JXS}
J_{XS}(k)
&=&\left(\frac{g_{\pi N}}{2m_N}\right)^2\tau_1\cdot\tau_2\int dr \frac{3  m_\pi^3 }{4 \sqrt{2} \pi  k m_N} j_1\left(\frac{k r}{2}\right)\left(u w' -u'w\right)Y_2
\eea
\end{subequations}
The $t^{ij}$ contribution can be parametrized as ~\cite{Polyakov:2019lbq}
\bea
\label{SEAIJ}
&&\hspace*{-0.5cm}t^{ij}=\left<+\frac{k}{2} m'\bigg|T_{XS}^{ij}\bigg|-\frac{k}{2} m\right>
\nonumber\\
&=&\frac{k^ik^j-\delta^{ij}k^2}{4m_D} D_{0,XS}\delta_{m'm}
+\frac{(k^jk^\alpha Q_{m'm}^{i\alpha}+k^ik^\alpha Q_{m'm}^{j\alpha}-k^2Q_{m'm}^{ij}-\delta^{ij}Q_{m'm}^{\alpha\beta}k_\alpha k_\beta)}{2m_D}D_{2,XS}
\nonumber\\
&+&\frac{1}{4m_D}(k^ik^j-\delta^{ij}k^2)Q_{m'm}^{\alpha\beta}\hat{k}_\alpha \hat{k}_\beta D_{3,XS}+Q_{m'm}^{ij} D_{4,XS} + \delta^{ij}\delta_{m'm} D_{5,XS}
+\delta^{ij}\delta_{m'm}D_{6,XS}Q^{\alpha\beta}\hat{k}_\alpha \hat{k} _\beta,\nonumber\\
\eea
where $D_{0,XS}$ , $D_{2,XS}$ and $D_{3,XS}$ are the form factors which satisfy the current conserved condition.
The contributions $\delta^{ij}$ and $Q^{ij}$ in (\ref{SEAIJ}), appear to upset the conservation law $k^iT^{ij}$. As we show below, the conservation law is upheld in the non-relativistic limit when all 
contributions are added up. As a result, the contributions $\delta^{ij}$ and $Q^{ij}$ cancel out. With this in mind, we can
extract them by defining
\bea
\label{eq:tijnew}
\tilde{t}^{ij}=t^{ij}-\frac{1}{3}\delta^{ij}t^{l}_l-\frac{3}{5}Q^{ij}t^{\alpha\beta}Q^{\alpha\beta}
\eea
which  is traceless and satisfies $\tilde{t}^{ij}Q^{ij}=0$. 
Using $\tilde{t}^{ij}$, one can solve for the form factors $D_{\{0,2,3\},XS}$ as
\begin{subequations}
\bea
\label{eq:FF_D0XS}
D_{0,XS}&=&\frac{g_{\pi N}^2}{(2m_N)^2}\tau_1.\tau_2m_\pi^2 \int dr j_2\left(\frac{k r}{2}\right) \frac{2m_D\left(-m_\pi r u^2 Y_2+2 \sqrt{2} u w (3\bar{Y}_1-2 m_\pi r Y_2)+w^2 (m_\pi r Y_2-3\bar{Y}_1)\right)}{3 \pi  k^2 r}
\nonumber\\
\\
\label{eq:FF_D2XS}
D_{2,XS}&=&-\frac{g_{\pi N}^2\tau_1.\tau_2}{(2m_N)^2}\frac{m_Dm_\pi^2 }{14 \pi  k^2 }\int \frac{dr}{r}\Bigg\{j_2\left(\frac{k r}{2}\right) \left[14 m_\pi r u^2 Y_2+7 \sqrt{2} u w (6 \bar{Y}_1-m_\pi r Y_2)+2 w^2 (2 m_\pi r Y_2+3 \bar{Y}_1)\right]
\nonumber\\
&+&18 w^2 j_4\left(\frac{k r}{2}\right) (m_\pi r Y_2-2 \bar{Y}_1)\Bigg\}
\\
\label{eq:FF_D3XS}
D_{3,XS}&=&\frac{g_{\pi N}^2}{(2m_N)^2}\tau_1.\tau_2\frac{9 m_D m_\pi^2 w^2 }{\pi  k^2}\int \frac{dr}{r} j_4\left(\frac{k r}{2}\right) (m_\pi r Y_2-2 \bar{Y}_1)
\eea
\end{subequations}
\end{widetext}

\begin{widetext}
\subsection{Fig.~\ref{fig:NR_diagrams}c : Pion exchange contribution}
The non relativistic reduction of the pion exchange contribution in Fig.~\ref{fig:NR_diagrams}c,  yields
\bea\label{eq:tmunupion_off}
T_{X\pi\pi}^{00}&=&-\frac{g_{\pi N}^2}{(2m_N)^2}\tau_1\cdot\tau_2\frac{\vec{k}^2 \vec{\sigma_1}.\vec{q_1} \vec{\sigma_2}.\vec{q_2}}{2 w_{q_1}^2 w_{q_2}^2}T_{1\pi}(k)+\frac{g_{\pi N}^2}{(2m_N)^2}\tau_1\cdot\tau_2\frac{\sigma_1.q_1 \sigma_2.q_2 }{2}\left(\frac{1}{w_{q_2}^2}+\frac{1}{w_{q_1}^2}\right)+{\cal O}\bigg(\frac{g_{\pi N}^2}{m_N^4}\bigg)
\nonumber\\
T_{X\pi\pi}^{0j}&=&-\frac{g_{\pi N}^2}{(2m_N)^2}\tau_1\cdot\tau_2\frac{(q_1^j +q_2^j)\vec{P}_d.\vec{k} \sigma_1.q_1 \sigma_2.q_2}{4m_N w_{q_1}^2 w_{q_2}^2}T_{2\pi}(k)+{\cal O}\bigg(\frac{g_{\pi N}^2}{m_N^4}\bigg)
\nonumber\\
T_{X\pi\pi}^{ij}&=&\frac{g_{\pi N}^2}{(2m_N)^2}\tau_1\cdot\tau_2\left(\frac{(\vec{k}^2\delta^{ij}-k^ik^j) \vec{\sigma_1}.\vec{q_1} \vec{\sigma_2}.\vec{q_2}}{2 w_{q_1}^2 w_{q_2}^2}T_{1\pi}(k)+\frac{(q_1^i+q_2^i)(q_1^j+q_2^j)\vec{\sigma_1}.\vec{q_1} \vec{\sigma_2}.\vec{q_2}}{2 w_{q_1}^2 w_{q_2}^2}T_{2\pi}(k)\right)
\nonumber\\
&-&\delta^{ij}\frac{g_{\pi N}^2}{(2m_N)^2}\tau_1\cdot\tau_2\frac{\sigma_1.q_1 \sigma_2.q_2  \left(\vec{q}_1^2+\vec{q}_2^2+2m_\pi^2\right)}{2 w_{q_1}^2 w_{q_2}^2}+{\cal O}\bigg(\frac{g_{\pi N}^2}{m_N^4}\bigg)
\eea
The second contributions in each of  $T_{X\pi\pi}^{00}$ and $T_{X\pi\pi}^{ij}$ follow from the off-shell part in~(\ref{eq:onshellpi}). Their contribution to the deuteron EMT, are
\bea
\label{eq:pipi}
\left<+\frac{k}{2} m'\bigg|T_{X\pi\pi}^{00}\bigg|-\frac{k}{2} m\right>&=&m_D A_{X\pi\pi}\delta_{mm'}+Q_{X\pi\pi}\frac{\langle m'|Q^{\alpha\beta}|m\rangle k^\alpha k^\beta}{2m_D}
\nonumber\\
\left<+\frac{k}{2} m'\bigg|T_{X\pi\pi}^{0j}\bigg|-\frac{k}{2} m\right>&=&\frac{\langle m'|(\vec{S}\times i\vec{k})^j|m\rangle}{2}J_{X\pi\pi}
\nonumber\\
\left<+\frac{k}{2} m'\bigg|T_{X\pi\pi}^{ij}\bigg|-\frac{k}{2} m\right>
&=&\frac{k^ik^j-\delta^{ij}k^2}{4m_D} D_{0,X\pi\pi}\delta_{m'm}
+\frac{(k^jk^\alpha Q_{m'm}^{i\alpha}+k^ik^\alpha Q_{m'm}^{j\alpha}-k^2Q_{m'm}^{ij}-\delta^{ij}Q_{m'm}^{\alpha\beta}k_\alpha k_\beta)}{2m_D}D_{2,X\pi\pi}
\nonumber\\
&+&\frac{1}{4m_D}(k^ik^j-\delta^{ij}k^2)Q_{m'm}^{\alpha\beta}\hat{k}_\alpha \hat{k}_\beta D_{3,X\pi\pi}
\eea
For the $T^{ij}$ component, we only keep the contributions that satisfy the current conservation law.  As we noted above, the upsetting terms cancel out when all contributions are added up. The corresponding form factors are lengthy and given in Appendix~\ref{APP_PIONX}. 
\end{widetext}

\begin{widetext}
\subsection{Fig.~\ref{fig:NR_diagrams}d : Recoil contribution}
For the recoil contribution in left panel in Fig.~\ref{fig:NR_diagrams}d, the amplitude can be written as
\bea
T_{Rc}^{\mu\nu}=\frac{g_{\pi N}^2}{(2m_N)^2}\frac{\bar{u}(p_1^\prime)\slashed{q_2}\gamma_5\tau_1^a(\gamma_0 E_p-\vec{\gamma}.(\vec{p_1}+\vec{k})+m)T^{\mu\nu}_Nu(p_1)}{4 E_p w_{q_2} (E_p-E_{p_1^\prime}+w_{q_2}) (E_{p_2^\prime}-E_{p_2}+w_{q_2}))}\bar{u}(p_2')\slashed{q}_2\gamma_5\tau_2^a u(p_2)
\eea
As for the disconnected diagrams, which contribute to the wave function renormalization factor, they can be written as~\cite{Gari:1976kj}
\bea
T_{wf}^{\mu\nu}=-\frac{1}{2}T^{\mu\nu}_{Rc}-\frac{g_{\pi N}^2}{(2m_N)^2}\frac{\bar{u}(p_1^\prime)\slashed{q_2}\gamma_5\tau_1^a(\gamma_0 E_p-\vec{\gamma}.(\vec{p_1}+\vec{k})+m)T^{\mu\nu}_Nu(p_1)}{8 E_p w_{q_2} (E_{p_1^\prime}-E_p+w_{q_2})(E_{p_2}-E_{p_2^\prime}+w_{q_2}))}\bar{u}(p_2')\slashed{q}_2\gamma_5\tau_2^a u(p_2)
\eea
As a result, the recoil contribution and wave function renormalization $T^{\mu\nu}_{XR}=T^{\mu\nu}_{Rc}+T^{\mu\nu}_{wf}$ can be expressed as
\bea
\label{RECINT}
T^{00}_{XR}
&=&-\frac{g_{\pi N}^2}{(2m_N)^2}\tau_1\cdot\tau_2A(k)\frac{q_2.k \sigma_1.q_2 \sigma_2.q_2}{2 w_{q_2}^4}+(1\leftrightarrow 2)+{\cal O}\bigg(\frac {g_{\pi N}^2}{m_N^4}\bigg)
\nonumber\\
T^{0j}_{XR}&=&\frac{g_{\pi N}^2}{(2m_N)^2}\tau_1\cdot\tau_2\Bigg\{\left(-\frac{P_d^j k.q_2 \,\sigma_1.q_2 \sigma_2.q_2}{4m_N w_{q_2}^4}\right)A(k)+(A(k)+B(k))\left(\frac{i  k.q_2\, \sigma_2.q_2 (k\times q_2)^j}{8 m_Nw_{q_2}^4}\right)\Bigg\}
\nonumber\\
&+&(1\leftrightarrow 2)+{\cal O}\bigg(\frac {g_{\pi N}^2}{m_N^4}\bigg)
\nonumber\\
T^{ij}_{XR}&=&\bigg(\frac {g_{\pi N}^2}{m_N^4}\bigg)
\eea
The contribution to the GFF from Fig.~\ref{fig:NR_diagrams}d can be expressed as
\bea
\label{eq:XR}
\left<+\frac{k}{2} m'\bigg|T_{XR}^{00}\bigg|-\frac{k}{2} m\right>&=&m_D A_{XR}\delta_{mm'}+Q_{XR}\frac{\langle m'|Q^{\alpha\beta}|m\rangle k^\alpha k^\beta}{2m_D}
\nonumber\\
\left<+\frac{k}{2} m'\bigg|T_{XR}^{0j}\bigg|-\frac{k}{2} m\right>&=&\frac{\langle m'|(\vec{S}\times i\vec{k})^j|m\rangle}{2}J_{XR}
\eea
where 
\begin{subequations}
\bea
\label{eq:FF_AXR}
A_{XR}(k)
&=&\frac{g_{\pi N}^2\tau_1\cdot\tau_2A(k)}{(2m_N)^2m_D}\int dr \frac{k m_\pi^2 }{24 \pi }j_1\left(\frac{k r}{2}\right) \bigg[u^2 (m_\pi r Y_2-5 \bar{Y}_1)+4 \sqrt{2} u w (m_\pi r Y_2-2 \bar{Y}_1)
-w^2 (m_\pi r Y_2+\bar{Y}_1)\bigg]
\nonumber\\
\\
\label{eq:FF_QXR}
Q_{XR}(k)
&=&\frac{g_{\pi N}^2\tau_1\cdot\tau_2A(k)}{(2m_N)^2}m_D\int dr \frac{m_\pi^2}{40 \pi k} \Bigg\{j_1\left(\frac{k r}{2}\right) \bigg[8 u^2 (m_\pi r Y_2-5 \bar{Y}_1)
+4 \sqrt{2} u w (2 \bar{Y}_1-m_\pi r Y_2)
+10 w^2 (m_\pi r Y_2+\bar{Y}_1)\bigg]
\nonumber\\
&+&3 j_3\left(\frac{k r}{2}\right) \bigg[-4 m_\pi r u^2 Y_2+2 \sqrt{2} u w (m_\pi r Y_2+3 \bar{Y}_1)+5 w^2 (3 \bar{Y}_1-m_\pi r Y_2)\bigg]\Bigg\}
\\
\label{eq:FF_JXR}
J_{XR}(k)
&=&\left(\frac{g_{\pi N}}{2m_N}\right)^2A(k)\tau_1\cdot\tau_2\int dr \frac{3 m_\pi^2 }{16 \pi  k m_N r^2}\Bigg\{2 j_1\left(\frac{k r}{2}\right) \bigg[r \left(2 \sqrt{2} u (w' \bar{Y}_1-m_\pi w Y_2)+m_\pi w^2 Y_2-2 \sqrt{2} u' w \bar{Y}_1\right)
\nonumber\\
&+&6 w^2 \bar{Y}_1\bigg]-k r w j_2\left(\frac{k r}{2}\right) \bigg[(2 \sqrt{2} u (\bar{Y}_1-m_\pi r Y_2)+w (m_\pi r Y_2+2 \bar{Y}_1)\bigg]
\nonumber\\
&+&k r w \bar{Y}_1 \left(2 \sqrt{2} u-w\right) j_0\left(\frac{k r}{2}\right)\Bigg\}
\nonumber\\
&+&\left(\frac{g_{\pi N}}{2m_N}\right)^2\left(A(k)+B(k)\right)\tau_1\cdot\tau_2\int dr \frac{k m_\pi^2 }{160 \pi  m_N}\Bigg\{j_1\left(\frac{k r}{2}\right) \bigg[2 u^2 (m_\pi r Y_2-5 \bar{Y}_1)
\nonumber\\
&-&\sqrt{2} u w (\bar{Y}_1-2 m_\pi r Y_2)+w^2 (m_\pi r Y_2+4 \bar{Y}_1)\bigg]
\nonumber\\
&+&j_3\left(\frac{k r}{2}\right) \bigg[2 m_\pi r u^2 Y_2+2 \sqrt{2} u w (m_\pi r Y_2-3 \bar{Y}_1)+w^2 (m_\pi r Y_2-6 \bar{Y}_1)\bigg]\Bigg\}
\eea
\end{subequations}
\end{widetext}

\subsection{EMT conservation in non-relativistic limit}\label{sec:conser}
To check (\ref{NREMTX}) in the non-relativistic limit, we combine 
the result in (\ref{eq:tmunuKR}) with that in (\ref{eq:tmunupion_off}), 
to have
\begin{widetext}
\bea
k^iT^{ij}_{XS}+k^iT^{ij}_{X\pi\pi}&=&-\frac{g_{\pi N}^2}{(2m_N)^2}\tau_1\cdot\tau_2\left(\frac{\sigma_2.q_2k.q_2\sigma_1^j+\sigma_1.k\sigma_2.q_2q_2^j}{2w_{q_2}^2}+\frac{\sigma_1.q_1k.q_1\sigma_2^j+\sigma_2.k\sigma_1.q_1q_1^j}{2w_{q_1}^2}\right)
\nonumber\\
&+&\frac{g_{\pi N}^2}{(2m_N)^2}\tau_1\cdot\tau_2\left(\frac{q_1^j}{w_{q_1}^2}-\frac{q_2^j}{w_{q_2}^2}\right)\sigma_1.q_1\sigma_2.q_2
\nonumber\\
&=&\frac{g_{\pi N}^2\tau_1\cdot\tau_2}{(2m_N)^2}\left(\frac{-\sigma_2.q_2k.q_2\sigma_1^j+\sigma_1.(k-2q_2)\sigma_2.q_2q_2^j}{2w_{q_2}^2}-\frac{\sigma_1.q_1k.q_1\sigma_2^j+\sigma_2.(k-2q_2)\sigma_1.q_1q_1^j}{2w_{q_1}^2}\right)
\nonumber\\
\eea
where we have set $T_{1\pi}=T_{2\pi}=1$.
On the other hand, $[V_\pi, T_N^{0\nu}]$ can be expressed as
\bea
[V_\pi,T^{0j}]&=&-\frac{g_{\pi N}^2\tau_1\cdot\tau_2}{(2m_N)^2}\left(\frac{\sigma_1.q_2\sigma_2.q_2}{w_{q_2}^2}(p_1^j+\frac{k^j}{2}+\frac{(\sigma_1\times ik)^j}{4})-(p_1'^j-\frac{k^j}{2}+\frac{(\sigma_1\times ik)^j}{4})\frac{\sigma_1.q_2\sigma_2.q_2}{w_{q_2}^2}\right)
\nonumber\\
&-&\frac{g_{\pi N}^2\tau_1\cdot\tau_2}{(2m_N)^2}\left(\frac{\sigma_1.q_1\sigma_2.q_1}{w_{q_1}^2}(p_2^j+\frac{k^j}{2}+\frac{(\sigma_2\times ik)^j}{4})-(p_2'^j-\frac{k^j}{2}+\frac{(\sigma_2\times ik)^j}{4})\frac{\sigma_1.q_1\sigma_2.q_1}{w_{q_1}^2}\right)
\nonumber\\
&=&-\frac{g_{\pi N}^2\tau_1\cdot\tau_2}{(2m_N)^2}\left(\frac{\sigma_2.q_2k.q_2\sigma_1^j-\sigma_1.(k-2q_2)\sigma_2.q_2q_2^j}{2w_{q_2}^2}+\frac{\sigma_1.q_1k.q_1\sigma_2^j+\sigma_2.(k-2q_2)\sigma_1.q_1q_1^j}{2w_{q_1}^2}\right)
\eea
where we have set $A(k^2)=1$ and $B(k^2)=0$, which are the leading  ChPT values.
\end{widetext}

\begin{figure*}[htbp] 
\begin{minipage}[b]{.45\linewidth}
\hspace*{-0.3cm}\includegraphics[width=1.1\textwidth, height=5.5cm]{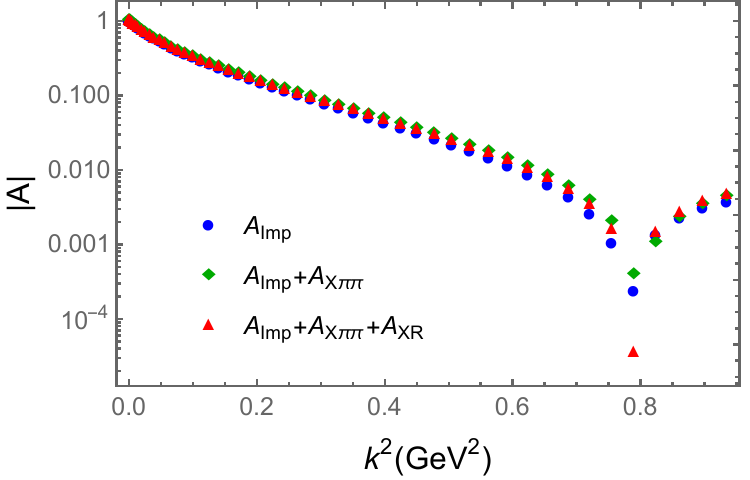}   
 \vspace{0pt}
\end{minipage}
\hfill
\begin{minipage}[b]{.45\linewidth}   
\hspace*{-0.55cm} \includegraphics[width=1.1\textwidth, height=5.5cm]{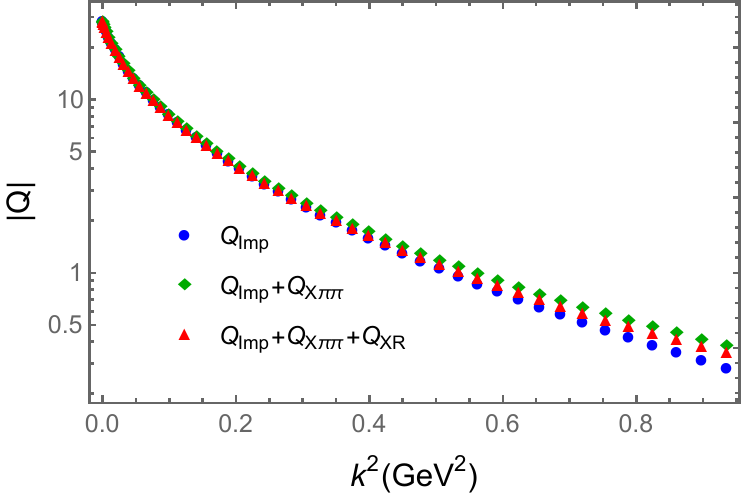} 
   \vspace{0pt}
\end{minipage}  
\\[-0.4cm]
\begin{minipage}[t]{.45\linewidth}
\hspace*{-0.5cm}\includegraphics[width=1.12\textwidth, height=5.5cm]{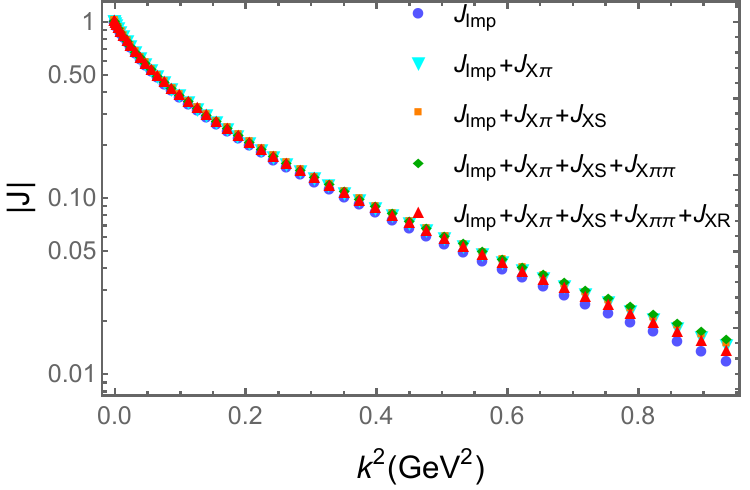}   
 \vspace{0pt}
\end{minipage}
\hfill
\begin{minipage}[t]{.45\linewidth}   
\hspace*{-0.85cm} \includegraphics[width=1.1\textwidth, height=5.5cm]{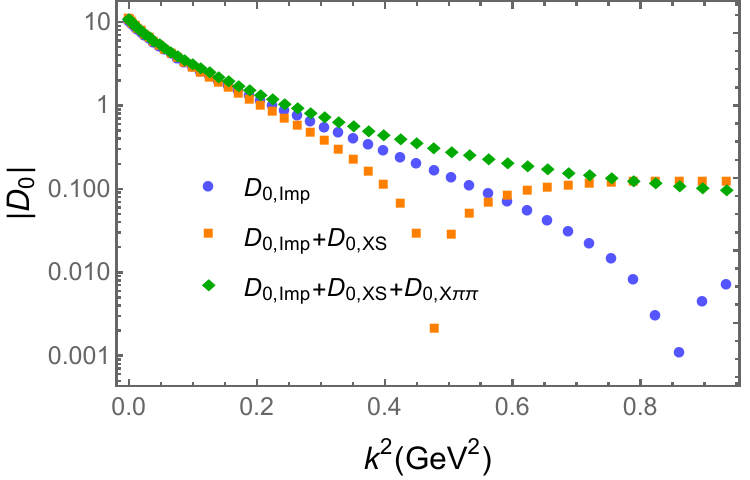}  
   \vspace{0pt}
\end{minipage} 
\begin{minipage}[t]{.45\linewidth}
\hspace*{-0.5cm}\includegraphics[width=1.12\textwidth, height=5.5cm]{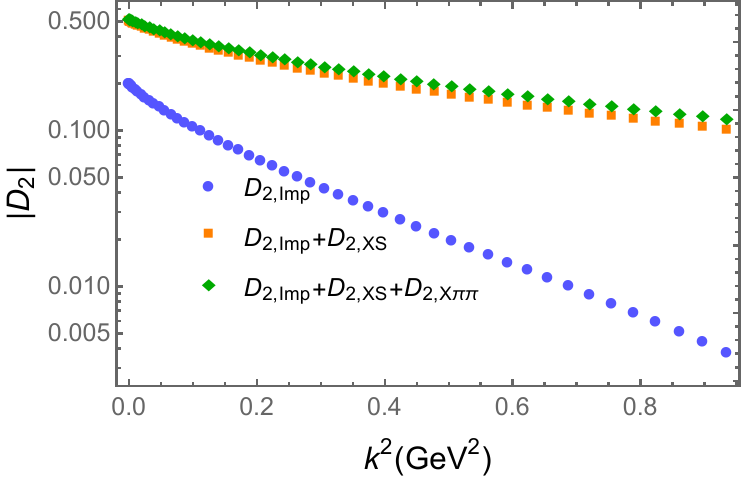}   
 \vspace{0pt}
\end{minipage}
\hfill
\begin{minipage}[t]{.45\linewidth}   
\hspace*{-0.85cm} \includegraphics[width=1.1\textwidth, height=5.5cm]{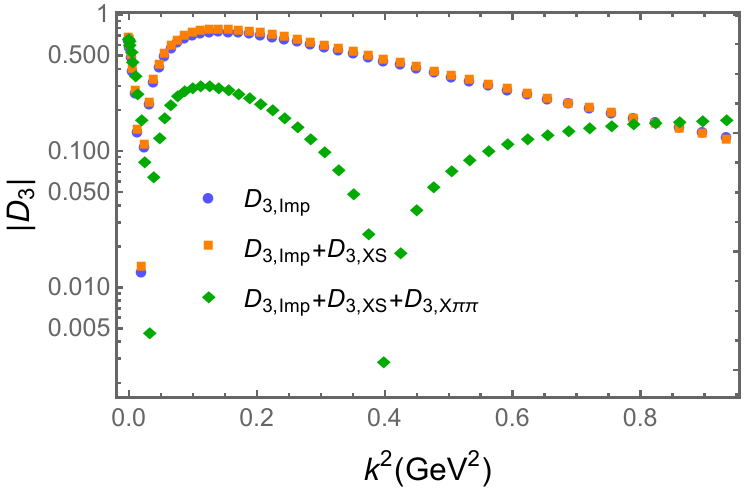}  
   \vspace{-12pt}
\end{minipage}
\caption{The deuteron GFFs in the impulse approximation from~\cite{He:2023ogg},
along with  exchange corrections from this work. The subscript ``imp" refer to the impulse approximation, while the subsripts  ``$X\pi$", ``$XS$", ``$X\pi\pi$" and ``XR" refer to the exchange contributions from 
the pair diagram~(Fig.~\ref{fig:NR_diagrams}a),  seagull diagram~(Fig.~\ref{fig:NR_diagrams}b), pion exchange diagram~(Fig.~\ref{fig:NR_diagrams}c) and recoil contribution~(Fig.~\ref{fig:NR_diagrams}d), respectively.
These results are for the pion GFFs set to $T_{1\pi}=T_{2\pi}=1$.}
\label{fig:ex_result_tree}
\end{figure*}

\begin{figure*}[htbp] 
\begin{minipage}[b]{.45\linewidth}
\hspace*{-0.3cm}\includegraphics[width=1.1\textwidth, height=5.5cm]{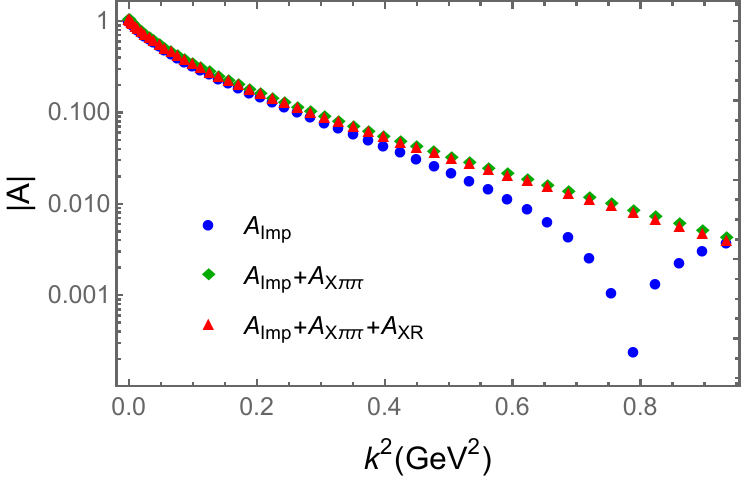}   
 \vspace{0pt}
\end{minipage}
\hfill
\begin{minipage}[b]{.45\linewidth}   
\hspace*{-0.55cm} \includegraphics[width=1.1\textwidth, height=5.5cm]{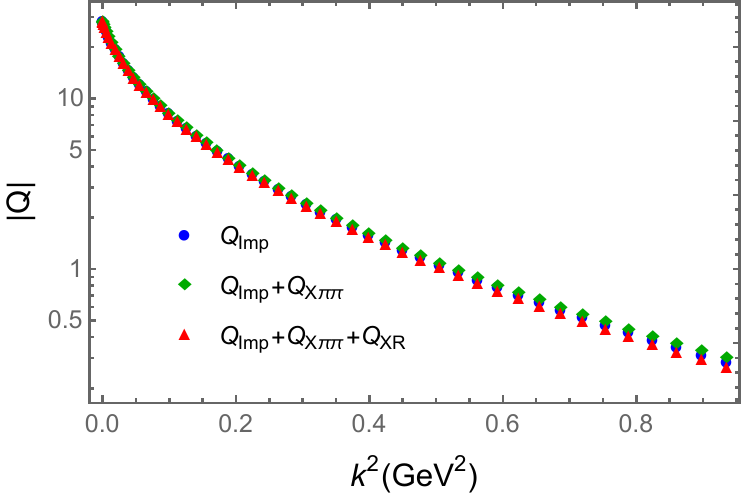} 
   \vspace{0pt}
\end{minipage}  
\\[-0.4cm]
\begin{minipage}[t]{.45\linewidth}
\hspace*{-0.5cm}\includegraphics[width=1.12\textwidth, height=5.5cm]{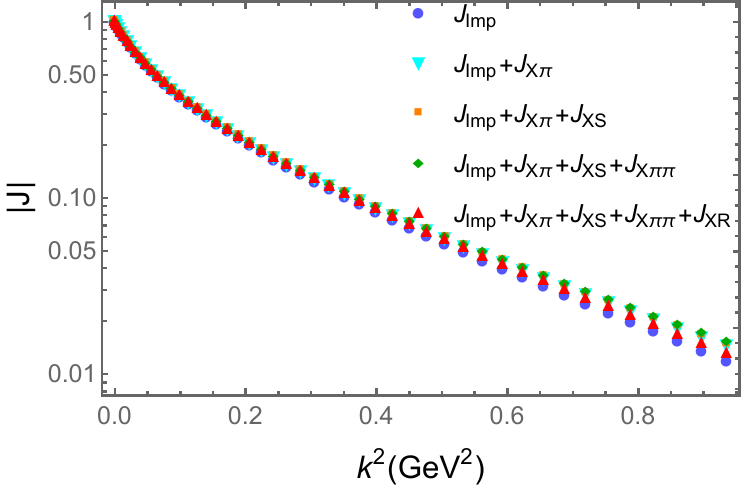}   
 \vspace{0pt}
\end{minipage}
\hfill
\begin{minipage}[t]{.45\linewidth}   
\hspace*{-0.85cm} \includegraphics[width=1.1\textwidth, height=5.5cm]{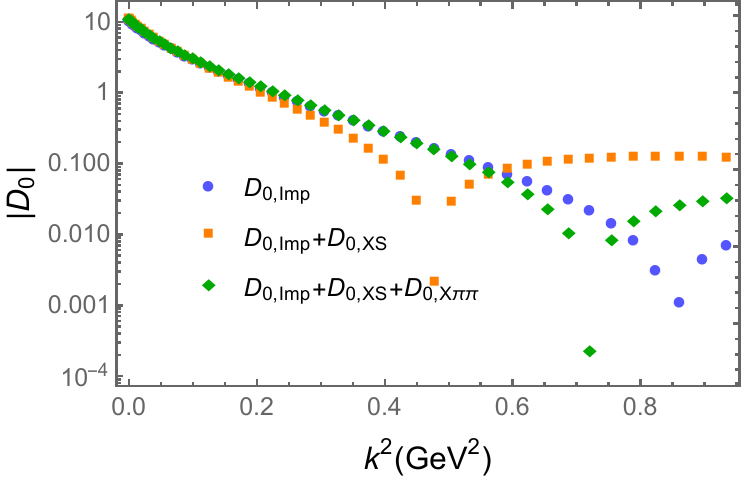}  
   \vspace{0pt}
\end{minipage} 
\begin{minipage}[t]{.45\linewidth}
\hspace*{-0.5cm}\includegraphics[width=1.12\textwidth, height=5.5cm]{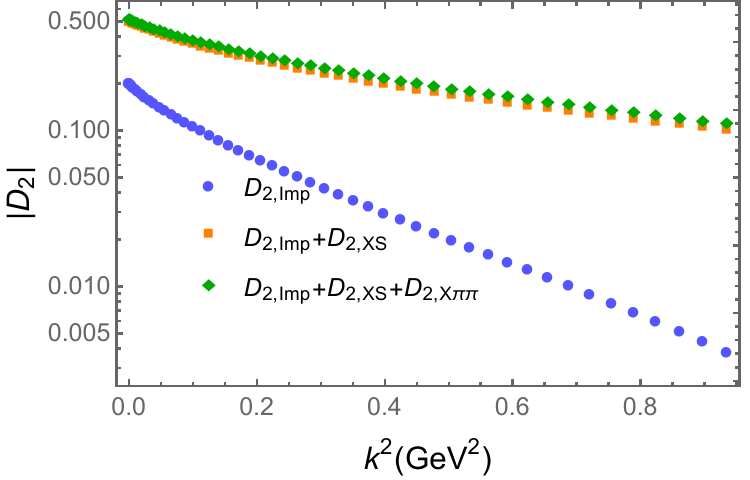}   
 \vspace{0pt}
\end{minipage}
\hfill
\begin{minipage}[t]{.45\linewidth}   
\hspace*{-0.85cm} \includegraphics[width=1.1\textwidth, height=5.5cm]{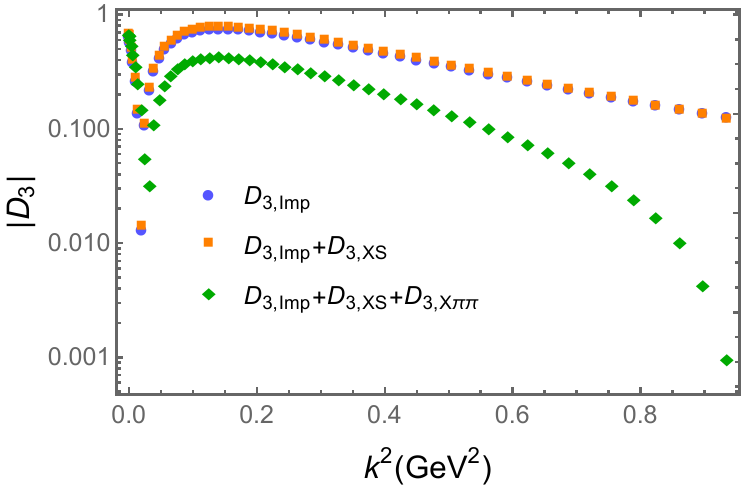}  
   \vspace{-12pt}
\end{minipage}
\caption{Same as Fig.~\ref{fig:ex_result_tree} but with the parametrized pion GFFs following from the QCD Lattice calculation in~\cite{Hackett:2023nkr} in Eq.~\ref{eq:GFFpiLattice}.}
\label{fig:ex_result}
\end{figure*}

\section{Numerical results}
\label{SEC_IV}
To summarise, the exchange contributions in the deuteron state are 
\bea
\label{SUMDX}
T^{00}_{DX}&=&T_{X\pi\pi}^{00}+T^{00}_{XR}\nonumber\\
T^{0j}_{DX}&=&T^{0j}_{X\pi}+T_{XS}^{0j}+T^{0j}_{X\pi\pi}+T^{0j}_{XR}\nonumber\\
T^{ij}_{DX}&=&T_{XS}^{ij}+T^{ij}_{X\pi\pi}
\eea
The corresponding GFFs from the exchange currents are
\bea\label{eq:FF_sum}
&&A_X=A_{X\pi\pi}(\ref{eq:FF_AXpipi})+A_{XR}(\ref{eq:FF_AXR})\nonumber\\
&&Q_X=Q_{X\pi\pi}(\ref{eq:FF_QXpipi})+Q_{XR}(\ref{eq:FF_QXR})\nonumber\\
\hspace*{-8cm}&&J_X=J_{X\pi}(\ref{eq:FF_JXpi})+J_{XS}(\ref{eq:FF_JXS})+J_{X\pi\pi}(\ref{eq:FF_JXpipi})+J_{XR}(\ref{eq:FF_JXR})\nonumber\\
&&D_{0,X}=D_{0,XS}(\ref{eq:FF_D0XS})+D_{0,X\pi\pi}(\ref{eq:FF_D0Xpipi})\nonumber\\
&&D_{2,X}=D_{3,XS}(\ref{eq:FF_D2XS})+D_{3,X\pi\pi}(\ref{eq:FF_D2Xpipi})\nonumber\\
&&D_{3,X}=D_{3,XS}(\ref{eq:FF_D3XS})+D_{3,X\pi\pi}(\ref{eq:FF_D3Xpipi})
\eea
with the contributions from the a-pair, b-seagull, c-pipi and d-recoil
labeled by the pertinent result.
In the forward limit, we note that $T^{00}_{DX}=-V_\pi$ which is  the pion pair potential.
When combined with  the deuteron result $T_D^{00}$  from the impulse approximation in~\cite{He:2023ogg}, we  obtain the sum rule for the deuteron mass 
$$m_D\sim 2m_N+\Bigg<\frac{P^2}{m_N^2}-V_\pi\Bigg>_D\,.$$
with the bracket including the recoil correction at next
to leading order in $P$, and the pion exchange correction. 
Note that the contribution of $-V_\pi$ to the deuteron energy from pion exchange through the mass GFF, is consistent with the one made in DIS in~\cite{Kaptari:1989un}. 
 We now proceed to compare the changes
due to (\ref{SUMDX}) on the results from our analysis, using the impulse approximation in~\cite{He:2023ogg}.

Since the conservation law in~\ref{sec:conser} is only satisfied at the leading order in ChPT, we first set  the pion GFF to their ChPT values  $T_{1\pi}=T_{2\pi}=1$ in our numerical results. For the nucleon GFFs, we use the ones detailed in our previous work~\cite{He:2023ogg}, which are borrowed from the Lattice calculation in~\cite{Hackett:2023rif} and the dual gravity calculation in~\cite{Mamo:2022eui}, for a direct comparison between the exchange currents and the impulse approximation. In Fig~\ref{fig:ex_result_tree} we show the results solely from the impulse approximation derived in~\cite{He:2023ogg}, in blue-dotted lines for all the deuteron GFFs.
The added  exchange contributions from the
pair, seagull, pipi and recoil as detailed in (\ref{eq:FF_sum}) for each of the deuteron GFFs are shown in cyan-, orange-, green-, red-dotted lines. 
To show the correction due to the  pion GFFs, we  borrow their parameterization from the recent Lattice calculation in~\cite{Hackett:2023nkr},
\bea\label{eq:GFFpiLattice}
T_{1\pi}^g(k)=\frac{0.596}{1+\frac{\vec{k}^2}{0.677~\text{GeV}^2}} \nonumber\\
T_{1\pi}^q(k)=\frac{0.304}{1+\frac{\vec{k}^2}{1.44~\text{GeV}^2}} \nonumber\\
T_{2\pi}^g(k)=\frac{0.546}{1+\frac{\vec{k}^2}{1.129~\text{GeV}^2}} \nonumber\\
T_{2\pi}^q(k)=\frac{0.481}{1+\frac{\vec{k}^2}{1.262~\text{GeV}^2}} 
\eea
The same results with the pion GFFs following the parametrization (\ref{eq:GFFpiLattice}) from the lattice results in~\cite{Hackett:2023nkr}, are shown in  Fig~\ref{fig:ex_result}. For the seagull and off-shellness part, we do not have the parameterization form currently, so they are only quoted from the leading order in ChPT. The main differences  are seen in the deuteron GFFs $|A, D_0, D_3|$, offering the possibility of extracting the pion GFFs from the deuteron exchange pion contribution, as we detail further below. 

The deuteron mass GFF  |A|  receives contributions solely from the impulse approximation (blue-circles)  and the pion exchange diagram (green-diamonds). There is no  seagull contribution, and the recoil contribution is very small (red-triangle). A comparison of Fig~\ref{fig:ex_result_tree}  
(without the pion form factors) and Fig~\ref{fig:ex_result} (with the pion form factors),
shows large deviations in the range $k>\frac 12 m_N$. 
This suggests that the pion GFF $T_{1\pi}$ can be extracted from a measurement of the deuteron mass GFF
$|A|$.
The  recoil corrections are small (red-triangle) due to a large cancellation between the recoil contribution and wave-function renormalization, much like   for the electromagnetic form factors~\cite{Gari:1976kj,Ohta:1975qv}.

The exchange contributions to the deuteron spin and quadrupole GFFs  $|J, Q|$ shown in~Fig~\ref{fig:ex_result}, are relatively small for  a wide range of $k\leq m_N$. 
The deuteron GFFs $|D_{0,2,3}|$ shown in~Fig~\ref{fig:ex_result}, receive exchange contributions from the seagull and pion ($X\pi\pi$)  contributions, but none from the recoil corrections. 

The correction from the seagull
plus pion exchange contributions for $|D_0|$ (standard D-term) in~Fig~\ref{fig:ex_result}, are small for $k\leq \frac 12 m_N$. The seagull contribution moves down the diffractive peak from the impulse approximation (blue-circles) from about $0.9\,{\rm GeV}^2$ to about $0.5\,{\rm GeV}^2$ (orange-squares), before being shifted up (green-diamonds)  to about $0.7\,{\rm GeV}^2$ by the pion exchange contribution ($X\pi\pi$).

The exchange seagull contribution (orange-squares) is relatively large for $|D_2|$  (D-term for mixed tensor-spin-spin contribution) (orange-squares) in~Fig~\ref{fig:ex_result}, 
but very small for $|D_3|$ (D-term for mixed tensor-quadrupole contribution) for most of the range of $k$ supported by our approximation. This is mostly due to the fact that $D_2$ in the impulse approximation, is generated by a non-local interaction proportional to the D-wave admixture $w$ which is small, while the seagull exchange contribution (\ref{eq:FF_D2XS}) includes the contribution from the S-wavefunction $u$ which is large.

In contrast, the exchange pion contribution (green-diamonds) in~Fig~\ref{fig:ex_result},
shifts down $|D_3|$ at larger momenta. This observation once combined 
with the one for $|A|$ above, suggests that the  pion GFFs $T_{1\pi,2\pi}$ can be extracted from the exchange contributions in $|A,D_3|$.
With the exception of $|D_2|$, most of the GFFs mass, quadrupole and spin radii are left unchanged by the exchange corrections. We note that the pion exchange contribution in the standard deuteron D-term $D_0$ at large $k$, will modify the deuteron pressure distribution obtained in the impulse approximation at small distances.

\section{Conclusions}
\label{SEC_CON}
We have analyzed the exchange contributions to the deuteron EMT, by including the contributions from the pair, seagull, pion exchange and
recoil in the  non-relativistic Born approximation. This reduction is
expected to hold for momenta within a nucleon mass range. The exchange contributions were shown to satisfy the conservation law for the EMT.

The pion exchange contribution appears to be sizable in the deuteron  GFFs $|A,D_3|$ when the pion GFF are
added to the impulse approximation, with the seagull 
and recoil contributions being small in the range of
a nucleon mass. This suggests the possibility of empirically extracting the  pion GFFs from the measured deuteron GFFs $|A,D_3|$.

Most of the  exchange contributions are small in the range $k\leq \frac 12 m_N$, leaving unchanged the mass, quadrupole and spin radii in the deuteron. This may reflect on the fact that the low gravitational multipole transition rates maybe insensitive to the meson exchange currents, the exception being the deuteron $D_2$ GFF as we have explained above. This result may be  viewed as the gravitational analogue of Siegert theorem for the electromagnetic current in nuclei~\cite{Siegert:1937yt,Austern:1951zz}. However, this observation  deserves a more thorough investigation.

\vskip 0.5cm
{\noindent\bf Acknowledgements}

\noindent 
We thank Zein-Eddine Meziani for discussions.
This work is supported by the Office of Science, U.S. Department of Energy under Contract  No. DE-FG-88ER40388. This research is also supported in part within the framework of the Quark-Gluon Tomography (QGT) Topical Collaboration, under contract no. DE-SC0023646.

\appendix

\section{Pion exchange contribution}
\label{APP_PIONX}
The results of the GFFs for the pion exchange contribution in Fig.~\ref{fig:NR_diagrams}c are
\begin{widetext}
\begin{subequations}
\bea
\label{eq:FF_AXpipi}
A_{X\pi\pi}(k)&=&\frac{g_{\pi N}^2\tau_1\cdot\tau_2}{(2m_N)^2}T_{1\pi}(k)\int_{-1/2}^{1/2} dt\int dr \Bigg\{-\frac{k^2 }{1536 \pi m_D}j_0(k r t) \bigg[k^2 r \left(4 t^2-1\right) Y_0^{\pi\pi} \left(8 u^2-4 \sqrt{2} u w-9 w^2\right)
\nonumber\\
&-&72 L_\pi^2 r w \bar{Y}_1^{\pi\pi} \left(2 \sqrt{2} u-w\right)+24 L_\pi Y_0^{\pi\pi} \left(4 u^2-2 \sqrt{2} u w+5 w^2\right)\bigg]
\nonumber\\
&+&\frac{k j_1(k r t)}{768 \pi  m_D r t} \bigg[k^2 r Y_0^{\pi\pi} \left(32 L_\pi r t^2 \left(u^2+4 \sqrt{2} u w-w^2\right)-3 \left(4 t^2-1\right) w \left(2 \sqrt{2} u-7 w\right)\right)
\nonumber\\
&+&12 L_\pi \left(4 L_\pi r u^2 \bar{Y}_1^{\pi\pi}-2 \sqrt{2} u w (L_\pi r \bar{Y}_1^{\pi\pi}+3 Y_0^{\pi\pi})+w^2 (5 L_\pi r \bar{Y}_1^{\pi\pi}+3 Y_0^{\pi\pi})\right)\bigg]
\nonumber\\
&+&\frac{k^2 j_2(k r t)}{10752 \pi  m_D r t^2} \bigg[\left(4 t^2-1\right) w Y_0^{\pi\pi} \left(4 k^2 r^2 t^2 \left(63 \sqrt{2} u+25 w\right)-3255 w\right)
\nonumber\\
&-&56 L_\pi^2 r^2 t^2 \bar{Y}_1^{\pi\pi} \left(4 u^2-2 \sqrt{2} u w+5 w^2\right)+168 L_\pi r t^2 w Y_0^{\pi\pi} \left(2 \sqrt{2} u-w\right)\bigg]
\nonumber\\
&+&
j_4(k r t)\frac{31 k^4 r \left(4 t^2-1\right) w^2 Y_0^{\pi\pi}}{3584 \pi  m_D}
\Bigg\}
\nonumber\\
&+&\frac{g_{\pi N}^2\tau_1\cdot\tau_2}{(2m_N)^2}\int dr\frac{m_\pi^2 }{48 \pi  k m_D r^2}\Bigg\{-2 j_1\left(\frac{k r}{2}\right) \bigg[2 k^2 r^2 \bar{Y}_1 \left(u^2+4 \sqrt{2} u w-w^2\right)+12 m_\pi r u^2 Y_2
\nonumber\\
&-&6 \sqrt{2} u w (m_\pi r Y_2+3 \bar{Y}_1)+3 w^2 (5 m_\pi r Y_2+3 \bar{Y}_1)\bigg]
\nonumber\\
&-&3 k r j_0\left(\frac{k r}{2}\right) \bigg[2 \sqrt{2} u w (3 m_\pi r Y_2+\bar{Y}_1)-w^2 (3 m_\pi r Y_2+5 \bar{Y}_1)-4 u^2 \bar{Y}_1\bigg]
\nonumber\\
&+&k r j_2\left(\frac{k r}{2}\right) \bigg[4 m_\pi r u^2 Y_2-2 \sqrt{2} u w (m_\pi r Y_2+3 \bar{Y}_1)+w^2 (5 m_\pi r Y_2+3 \bar{Y}_1)\bigg]\Bigg\}
\\
\label{eq:FF_QXpipi}
Q_{X\pi\pi}(k)&=&\frac{g_{\pi N}^2\tau_1\cdot\tau_2}{(2m_N)^2}T_{1\pi}(k)\int_{-1/2}^{1/2} dt\int dr\Bigg\{-\frac{k^2 m_D rj_0(k r t)}{32 \pi } \bigg[\left(4 t^2-1\right) Y_0^{\pi\pi} \left(2 u^2-\sqrt{2} u w-3 w^2\right)\bigg] 
\nonumber\\
&+&\frac{k m_D Y_0^{\pi\pi} j_1(k r t)}{320 \pi  t} \bigg[16 L_\pi r t^2 \left(10 u^2-5 \sqrt{2} u w-w^2\right)-15 \left(4 t^2-1\right) w \left(2 \sqrt{2} u-7 w\right)\bigg]
\nonumber\\
&+&\frac{m_D j_2(k r t)}{448 \pi  r t^2} \bigg[\left(4 t^2-1\right) w^2 Y_0^{\pi\pi} \left(74 k^2 r^2 t^2-1407\right)-28 L_\pi^2 r^2 t^2 \bar{Y}_1^{\pi\pi} \left(4 u^2-2 \sqrt{2} u w+5 w^2\right)
\nonumber\\
&+&84 L_\pi r t^2 w Y_0^{\pi\pi} \left(2 \sqrt{2} u-w\right)\bigg]
\nonumber\\
&-&\frac{27 k m_D L_\pi r t w^2 \text{Y0} j_3(k r t)}{40 \pi }-\frac{3 k^2 m_D r \left(4 t^2-1\right) w^2 Y_0^{\pi\pi} j_4(k r t)}{448 \pi }\Bigg\}
\nonumber\\
&+&\frac{g_{\pi N}^2\tau_1\cdot\tau_2}{(2m_N)^2}\int dr \frac{m_D m_\pi^2 }{20 \pi  k^2 r}\Bigg\{-5 j_2\left(\frac{k r}{2}\right) \bigg[-4 m_\pi r u^2 Y_2+2 \sqrt{2} u w (m_\pi r Y_2+3 \bar{Y}_1)-w^2 (5 m_\pi r Y_2+3 \bar{Y}_1)\bigg]
\nonumber\\
&+&2 k r \bar{Y}_1 \left(-10 u^2+5 \sqrt{2} u w+w^2\right) j_1\left(\frac{k r}{2}\right)+27 k r w^2 \bar{Y}_1 j_3\left(\frac{k r}{2}\right)\Bigg\}
\\
\label{eq:FF_JXpipi}
J_{X\pi\pi}(k)&=&\frac{g_{\pi N}^2\tau_1\cdot\tau_2}{(2m_N)^2m_N}T_{2\pi}(k)\int_{-1/2}^{1/2} dt\int dr \Bigg\{\frac{L_\pi j_0(k r t) }{384 \pi  r}\bigg[72 L_\pi w \bar{Y}_1^{\pi\pi} \left(w-2 \sqrt{2} u\right)
\nonumber\\
&-&k^2 r \left(4 t^2-1\right) Y_0^{\pi\pi} \left(\sqrt{2} u (2 w-r w')+w \left(\sqrt{2} r u'-w\right)\right)\bigg]
\nonumber\\
&-&\frac{3 L_\pi j_1(k r t)}{320 \pi  k  r^2 t} \bigg[k^2 r w \left(-8 \sqrt{2} L_\pi r^2 t^2 u' \bar{Y}_1^{\pi\pi}+8 r t^2 \left(L_\pi w \bar{Y}_1^{\pi\pi}-2 \sqrt{2} u' Y_0^{\pi\pi}\right)+5 \left(4 t^2-1\right) w Y_0^{\pi\pi}\right)
\nonumber\\
&+&2 \sqrt{2} u k^2 r\left(-8 L_\pi r t^2 w \bar{Y}_1^{\pi\pi}+4 r t^2 w' (L_\pi r \bar{Y}_1^{\pi\pi}+2 Y_0^{\pi\pi})+\left(4 t^2+5\right) w Y_0^{\pi\pi}\right)
\nonumber\\
&+&20 L_\pi \left(r \left(2 \sqrt{2} u (w' \bar{Y}_1^{\pi\pi}-L_\pi w Y_2^{\pi\pi})+L_\pi w^2 Y_2^{\pi\pi}-2 \sqrt{2} u' w \bar{Y}_1^{\pi\pi}\right)+6 w^2 \bar{Y}_1^{\pi\pi}\right)\bigg]
\nonumber\\
&+&\frac{L_\pi j_2(k r t)}{2688 \pi   r^2 t^2} \bigg[2\sqrt{2}u w \left(\left(4 t^2-1\right) Y_0^{\pi\pi} \left(26 k^2 r^2 t^2-147\right)-504 L_\pi r t^2 \bar{Y}_1^{\pi\pi}\right)
\nonumber\\
&+&\sqrt{2}r u w' \left(1008 L_\pi r t^2 \bar{Y}_1^{\pi\pi}-\left(4 t^2-1\right) Y_0^{\pi\pi} \left(26 k^2 r^2 t^2-147\right)\right)
\nonumber\\
&-&2 r^2 t^2 w\left(13 k^2 \left(4 t^2-1\right) w Y_0^{\pi\pi}+504 \sqrt{2} L_\pi u' \bar{Y}_1^{\pi\pi}\right)+26 \sqrt{2} k^2 r^3 \left(4 t^2-1\right) t^2w u' Y_0^{\pi\pi}
\nonumber\\
&-&21 r \left(7 \sqrt{2} \left(4 t^2-1\right) w u' Y_0^{\pi\pi}-24 L_\pi t^2 w^2\bar{Y}_1^{\pi\pi}\right)+147 \left(4 t^2-1\right) w^2 Y_0^{\pi\pi}\bigg]
\nonumber\\
&-&\frac{3 k L_\pi t j_3(k r t)}{40 \pi  r} \bigg[\sqrt{2} u (2 w-r w') (3 Y_0^{\pi\pi}-L_\pi r \bar{Y}_1^{\pi\pi})+r w \left(\sqrt{2} u' (3 Y_0^{\pi\pi}-L_\pi r \bar{Y}_1^{\pi\pi})+L_\pi w \bar{Y}_1^{\pi\pi}\right)\bigg]
\nonumber\\
&+&j_4(k r t) \frac{11 k^2 L_\pi\left(4 t^2-1\right) Y_0^{\pi\pi}}{896 \pi }\bigg[\sqrt{2} u (2 w-r w')+w \left(\sqrt{2} r u'-w\right)\bigg]
\Bigg\}\\
\label{eq:FF_D0Xpipi}
D_{0,X\pi\pi}(k)&=&m_D\frac{g_{\pi N}^2\tau_1\cdot\tau_2}{(2m_N)^2}T_{1\pi}(k)\int_{-1/2}^{1/2} dt\int dr\Bigg\{\frac{j_0(k r t)}{384 \pi } \bigg[-k^2 r\left(4 t^2-1\right) Y_0^{\pi\pi} \left(8 u^2-4 \sqrt{2} u w-9 w^2\right)
\nonumber\\
&+&72 L_\pi^2 r w \bar{Y}_1^{\pi\pi} \left(2 \sqrt{2} u-w\right)-24 L_\pi Y_0^{\pi\pi} \left(4 u^2-2 \sqrt{2} u w+5 w^2\right)\bigg]
\nonumber\\
&+&\frac{j_1(k r t)}{192 \pi  k r t} \bigg[k^2 r Y_0^{\pi\pi} \left(32 L_\pi r t^2 \left(u^2+4 \sqrt{2} u w-w^2\right)-3 \left(4 t^2-1\right) w \left(2 \sqrt{2} u-7 w\right)\right)
\nonumber\\
&+&12 L_\pi \left(4 L_\pi r u^2 \bar{Y}_1^{\pi\pi}-2 \sqrt{2} u w (L_\pi r \bar{Y}_1^{\pi\pi}+3 Y_0^{\pi\pi})+w^2 (5 L_\pi r \bar{Y}_1^{\pi\pi}+3 Y_0^{\pi\pi})\right)\bigg]
\nonumber\\
&+&\frac{j_2(k r t)}{2688 \pi  r t^2} \bigg[\left(4 t^2-1\right) w Y_0^{\pi\pi} \left(4 k^2 r^2 t^2 \left(63 \sqrt{2} u+25 w\right)-3255 w\right)
\nonumber\\
&-&56 L_\pi^2 r^2 t^2 \bar{Y}_1^{\pi\pi} \left(4 u^2-2 \sqrt{2} u w+5 w^2\right)+168 L_\pi r t^2 w Y_0^{\pi\pi} \left(2 \sqrt{2} u-w\right)\bigg]
\nonumber\\
&+&\frac{j_4(k r t)}{896 \pi }31 k^2  r \left(4 t^2-1\right) w^2 \bar{Y}_0^{\pi\pi}
\Bigg\}
\nonumber\\
&+&m_D\frac{g_{\pi N}^2\tau_1\cdot\tau_2}{(2m_N)^2}T_{2\pi}(k)\int_{-1/2}^{1/2} dt\int dr\Bigg\{-\frac{j_0(k r t)}{864 \pi  k^2 r} \bigg[12 L_\pi^2 w \bar{Y}_1^{\pi\pi} \Big(k^2 r^2 \left(\sqrt{2} \left(40 t^2+17\right) u
-\left(68 t^2+25\right) w\right)
\nonumber\\
&+&120 w\Big)-72 k^2 L_\pi r t^2 Y_0^{\pi\pi} \left(28 u^2-14 \sqrt{2} u w-3 w^2\right)-9 k^4 r^2 t^2 \left(4 t^2-1\right) Y_0^{\pi\pi} \left(8 u^2-4 \sqrt{2} u w-9 w^2\right)
\nonumber\\
&+&40 L_\pi^4 r^2 Y_0^{\pi\pi} \left(4 u^2-2 \sqrt{2} u w+5 w^2\right)+40 L_\pi^3 r \Big(4 u^2 (Y_0^{\pi\pi}+5 Y_2^{\pi\pi})-2 \sqrt{2} u w (Y_0^{\pi\pi}+8 Y_2^{\pi\pi})
\nonumber\\
&+&w^2 (5 Y_0^{\pi\pi}-8 Y_2^{\pi\pi})\Big)\bigg]
\nonumber\\
&-&\frac{t j_1(k r t)}{240 \pi  k r} \bigg[k^2 r Y_0^{\pi\pi} \Big(2 L_\pi r \left(20 \left(8 t^2-1\right) u^2+32 \sqrt{2} \left(14 t^2-1\right) u w+\left(59-316 t^2\right) w^2\right)
\nonumber\\
&-&15 \left(4 t^2-1\right) w \left(2 \sqrt{2} u-7 w\right)\Big)-4 L_\pi \Big(28 L_\pi r u^2 (2 L_\pi r Y_2^{\pi\pi}-15 \bar{Y}_1^{\pi\pi})
\nonumber\\
&+&2 \sqrt{2} u w (L_\pi r (112 L_\pi r Y_2^{\pi\pi}-321 \bar{Y}_1^{\pi\pi})+105 Y_0^{\pi\pi})+w^2 (L_\pi r (321 \bar{Y}_1^{\pi\pi}-116 L_\pi r Y_2^{\pi\pi})-465 Y_0^{\pi\pi})\Big)\bigg]
\nonumber\\
&-&\frac{j_2(k r t)}{24192 \pi  k^4 r^3 t^2} \bigg[32 L_\pi^3 r \Big(4 u^2 \left(Y_0^{\pi\pi} \left(113 k^2 r^2 t^2-525\right)+Y_2^{\pi\pi} \left(124 k^2 r^2 t^2-2625\right)\right)
\nonumber\\
&+&2 \sqrt{2} u w \left(Y_0^{\pi\pi} \left(454 k^2 r^2 t^2+525\right)+8 Y_2^{\pi\pi} \left(13 k^2 r^2 t^2+525\right)\right)-w^2 Y_0^{\pi\pi} \left(2 k^2 r^2 t^2+2625\right)
\nonumber\\
&-&8w^2 Y_2^{\pi\pi} \left(176 k^2 r^2 t^2-525\right)\Big)+32 L_\pi^4 r^2 Y_0^{\pi\pi} \Big(4 u^2 \left(113 k^2 r^2 t^2-525\right)+2 \sqrt{2} u w \left(454 k^2 r^2 t^2+525\right)
\nonumber\\
&-&w^2 \left(2 k^2 r^2 t^2+2625\right)\Big)-3 L_\pi^2 \bar{Y}_1^{\pi\pi} \Big(k^4 r^4 t^2 \left(672 \left(24 t^2-1\right) u^2+224 \sqrt{2} \left(262 t^2-19\right) u w+\left(5977-9604 t^2\right) w^2\right)
\nonumber\\
&-&48 k^2 r^2 w \left(7 \sqrt{2} \left(532 t^2-109\right) u+\left(2234 t^2+1085\right) w\right)+201600 w^2\Big)
\nonumber\\
&+&288 k^2 L_\pi r t^2 w Y_0^{\pi\pi} \left(k^2 r^2 t^2 \left(546 \sqrt{2} u+179 w\right)-6510 w\right)
\nonumber\\
&+&36 k^4 r^2 \left(4 t^2-1\right) t^2 w Y_0^{\pi\pi} \left(4 k^2 r^2 t^2 \left(63 \sqrt{2} u+25 w\right)-3255 w\right)\bigg]
\nonumber\\
&+&\frac{L_\pi j_3(k r t)}{80 \pi  k^3 r^2 t^3} \bigg[-2 k^2 r t^2 \Big(16 L_\pi^2 r^2 t^2 u^2 Y_2^{\pi\pi}+64 \sqrt{2} L_\pi r t^2 u w (L_\pi r Y_2^{\pi\pi}-3 \bar{Y}_1^{\pi\pi})
\nonumber\\
&+&w^2 \left(2 L_\pi r t^2 (27 L_\pi r Y_2^{\pi\pi}-92 \bar{Y}_1^{\pi\pi})+735 \left(4 t^2-1\right) Y_0^{\pi\pi}\right)\Big)+k^4 r^3 \left(4 t^2-1\right) t^4 w Y_0^{\pi\pi} \left(32 \sqrt{2} u+61 w\right)
\nonumber\\
&+&140 L_\pi w^2 \left(20 L_\pi r t^2 Y_2^{\pi\pi}+\left(3-92 t^2\right) \bar{Y}_1^{\pi\pi}\right)\bigg]
\nonumber\\
&-&\frac{j_4(k r t)}{88704 \pi  k^2 r t^2} \bigg[3 L_\pi^2 w \bar{Y}_1^{\pi\pi} \left(4 k^2 r^2 t^2 \left(308 \sqrt{2} \left(4 t^2-1\right) u+\left(7364 t^2-3887\right) w\right)+99 \left(4613-16148 t^2\right) w\right)
\nonumber\\
&+&792 k^2 L_\pi r \left(836 t^2-147\right) t^2 w^2 Y_0^{\pi\pi}+12276 k^4 r^2 \left(4 t^2-1\right) t^4 w^2 Y_0^{\pi\pi}+1760 L_\pi^4 r^2 t^2 Y_0^{\pi\pi} \left(4 u^2-2 \sqrt{2} u w+5 w^2\right)
\nonumber\\
&+&352 L_\pi^3 r t^2 \left(20 u^2 (Y_0^{\pi\pi}+5 Y_2^{\pi\pi})-10 \sqrt{2} u w (Y_0^{\pi\pi}+8 Y_2^{\pi\pi})+w^2 (25 Y_0^{\pi\pi}-481 Y_2^{\pi\pi})\right)\bigg]
\nonumber\\
&+&
\frac{L_\pi j_5(k r t)}{64 \pi  k t}w^2 \left(28 k^2 r \left(4 t^2-1\right) t^2 Y_0^{\pi\pi}+L_\pi \left(\left(380 t^2-47\right) \bar{Y}_1^{\pi\pi}-48 L_\pi r t^2 Y_2^{\pi\pi}\right)\right)
\nonumber\\
&-&\frac{303 L_\pi^2 r \left(4 t^2-1\right) w^2 \bar{Y}_1^{\pi\pi} j_6(k r t)}{1408 \pi}
\Bigg\}
\\
\label{eq:FF_D2Xpipi}
D_{2,X\pi\pi}(k)&=&m_D\frac{g_{\pi N}^2\tau_1\cdot\tau_2}{(2m_N)^2}T_{2\pi}(k)\int_{-1/2}^{1/2} dt\int dr\Bigg\{\frac{L_\pi j_0(k r t)}{288 \pi  k^2 r} \bigg[-9 L_\pi w^2 \bar{Y}_1^{\pi\pi} \left(k^2 r^2 \left(4 t^2+1\right)-16\right)
\nonumber\\
&+&288 k^2 r t^2 Y_0^{\pi\pi} \left(2 u^2-\sqrt{2} u w+3 w^2\right)+4 L_\pi^3 r^2 Y_0^{\pi\pi} \left(4 u^2-2 \sqrt{2} u w+5 w^2\right)
\nonumber\\
&+&4 L_\pi^2 r \left(4 u^2 (Y_0^{\pi\pi}+5 Y_2^{\pi\pi})-2 \sqrt{2} u w (Y_0^{\pi\pi}+8 Y_2^{\pi\pi})+w^2 (5 Y_0^{\pi\pi}-8 Y_2^{\pi\pi})\right)\bigg]
\nonumber\\
&+&\frac{ L_\pi t j_1(k r t)}{20 \pi  k r} \bigg[8 L_\pi r u^2 (2 L_\pi r Y_2^{\pi\pi}-15 \bar{Y}_1^{\pi\pi})+4 \sqrt{2} u w (L_\pi r (3 \bar{Y}_1^{\pi\pi}-2 L_\pi r Y_2^{\pi\pi})+15 Y_0^{\pi\pi})
\nonumber\\
&+&w^2 (L_\pi r (17 L_\pi r Y_2^{\pi\pi}+30 \bar{Y}_1^{\pi\pi})-210 Y_0^{\pi\pi})\bigg]
\nonumber\\
&-&\frac{L_\pi j_2(k r t)}{4032 \pi  k^4 r^3 t^2} \bigg[8 L_\pi^2 r \Big(4 u^2 \left(Y_0^{\pi\pi} \left(26 k^2 r^2 t^2+105\right)+Y_2^{\pi\pi} \left(525-122 k^2 r^2 t^2\right)\right)
\nonumber\\
&-&2 \sqrt{2} u w \left(Y_0^{\pi\pi} \left(26 k^2 r^2 t^2+105\right)-44 k^2 r^2 t^2 Y_2^{\pi\pi}+840 Y_2^{\pi\pi}\right)
\nonumber\\
&+&5 w^2 \left(Y_0^{\pi\pi} \left(26 k^2 r^2 t^2+105\right)+8 Y_2^{\pi\pi} \left(20 k^2 r^2 t^2-21\right)\right)\Big)+8 L_\pi^3 r^2 Y_0^{\pi\pi} \left(4 u^2-2 \sqrt{2} u w+5 w^2\right) \left(26 k^2 r^2 t^2+105\right)
\nonumber\\
&-&9 L_\pi w \bar{Y}_1^{\pi\pi} \left(2 k^2 r^2 \left(168 \sqrt{2} \left(8 t^2-1\right) u-\left(668 t^2+63\right) w\right)+k^4 r^4 \left(580 t^2-41\right) t^2 w-3360 w\right)
\nonumber\\
&-&288 k^2 r t^2 w^2 Y_0^{\pi\pi} \left(37 k^2 r^2 t^2+147\right)\bigg]
\nonumber\\
&+&\frac{L_\pi j_3(k r t)}{80 \pi  k^3 r^2 t^3} \bigg[2 k^2 r t^2 \Big(32 L_\pi^2 r^2 t^2 u^2 Y_2^{\pi\pi}-16 \sqrt{2} L_\pi r t^2 u w (L_\pi r Y_2^{\pi\pi}+6 \bar{Y}_1^{\pi\pi})
\nonumber\\
&+&w^2 \left(4 L_\pi r t^2 (L_\pi r Y_2^{\pi\pi}-15 \bar{Y}_1^{\pi\pi})+45 \left(4 t^2-1\right) Y_0^{\pi\pi}\right)\Big)+15 k^4 r^3 \left(4 t^2-1\right) t^4 w^2 Y_0^{\pi\pi}
\nonumber\\
&+&60 L_\pi w^2 \left(\left(20 t^2+6\right) \bar{Y}_1^{\pi\pi}-11 L_\pi r t^2 Y_2^{\pi\pi}\right)\bigg]
\nonumber\\
&+&\frac{L_\pi j_4(k r t)}{14784 \pi  k^2 r t^2} \bigg[9 L_\pi w^2 \bar{Y}_1^{\pi\pi} \left(1576 k^2 r^2 t^4+90 t^2 \left(k^2 r^2-286\right)+3003\right)+396 k^2 r t^2 \left(21-172 t^2\right) w^2 Y_0^{\pi\pi}
\nonumber\\
&-&968 L_\pi^3 r^2 t^2 Y_0^{\pi\pi} \left(4 u^2-2 \sqrt{2} u w+5 w^2\right)
\nonumber\\
&-&88 L_\pi^2 r t^2 \left(44 u^2 (Y_0^{\pi\pi}+5 Y_2^{\pi\pi})-22 \sqrt{2} u w (Y_0^{\pi\pi}+8 Y_2^{\pi\pi})+5 w^2 (11 Y_0^{\pi\pi}-68 Y_2^{\pi\pi})\right)\bigg]
\nonumber\\
&+&\frac{3 L_\pi w^2 j_5(k r t) \left(2 k^2 r \left(4 t^2-1\right) t^2 Y_0^{\pi\pi}+L_\pi \left(\left(52 t^2-5\right) \bar{Y}_1^{\pi\pi}-8 L_\pi r t^2 Y_2^{\pi\pi}\right)\right)}{32 \pi  k t}
\nonumber\\
&-&\frac{81 L_\pi^2 r \left(4 t^2-1\right) w^2 \bar{Y}_1^{\pi\pi} j_6(k r t)}{704 \pi }\Bigg\}
\\
\label{eq:FF_D3Xpipi}
D_{3,X\pi\pi}(k)&=&m_D\frac{g_{\pi N}^2\tau_1\cdot\tau_2}{(2m_N)^2}T_{1\pi}(k)\int_{-1/2}^{1/2} dt\int dr\Bigg\{-\frac{k^2 r \left(4 t^2-1\right) Y_0^{\pi\pi} \left(2 u^2-\sqrt{2} u w-3 w^2\right) j_0(k r t)}{16 \pi }
\nonumber\\
&+&\frac{k Y_0^{\pi\pi} j_1(k r t) \left(16 L_\pi r t^2 \left(10 u^2-5 \sqrt{2} u w-w^2\right)-15 \left(4 t^2-1\right) w \left(2 \sqrt{2} u-7 w\right)\right)}{160 \pi  t}
\nonumber\\
&+&\frac{j_2(k r t)}{224 \pi  r t^2} \bigg[\left(4 t^2-1\right) w^2 Y_0^{\pi\pi} \left(74 k^2 r^2 t^2-1407\right)-28 L_\pi^2 r^2 t^2 \bar{Y}_1^{\pi\pi} \left(4 u^2-2 \sqrt{2} u w+5 w^2\right)
\nonumber\\
&+&84 L_\pi r t^2 w Y_0^{\pi\pi} \left(2 \sqrt{2} u-w\right)\bigg]-\frac{27 k L_\pi r t w^2 Y_0^{\pi\pi} j_3(k r t)}{20 \pi }-\frac{3 k^2 r \left(4 t^2-1\right) w^2 Y_0^{\pi\pi} j_4(k r t)}{224 \pi }\Bigg\}
\nonumber\\
&+&m_D\frac{g_{\pi N}^2\tau_1\cdot\tau_2}{(2m_N)^2}T_{2\pi}(k)\int_{-1/2}^{1/2} dt\int dr\Bigg\{\frac{j_0(k r t) }{96 \pi  k^2 r}\bigg[L_\pi^2 w \bar{Y}_1^{\pi\pi} \left(k^2 r^2 \left(2 \sqrt{2} \left(1-4 t^2\right) u+\left(580 t^2+143\right) w\right)-1296 w\right)
\nonumber\\
&-&2304 k^2 L_\pi r t^2 w^2 Y_0^{\pi\pi}+24 k^4 r^2 \left(4 t^2-1\right) t^2 Y_0^{\pi\pi} \left(2 u^2-\sqrt{2} u w-3 w^2\right)-36 L_\pi^4 r^2 Y_0^{\pi\pi} \left(4 u^2-2 \sqrt{2} u w+5 w^2\right)
\nonumber\\
&-&36 L_\pi^3 r \left(4 u^2 (Y_0^{\pi\pi}+5 Y_2^{\pi\pi})-2 \sqrt{2} u w (Y_0^{\pi\pi}+8 Y_2^{\pi\pi})+w^2 (5 Y_0^{\pi\pi}-8 Y_2^{\pi\pi})\right)\bigg]
\nonumber\\
&-&\frac{t j_1(k r t)}{40 \pi  k} \bigg[k^2 Y_0^{\pi\pi} \Big(8 L_\pi r \left(5 \left(8 t^2-1\right) u^2+\sqrt{2} \left(1-14 t^2\right) u w+\left(17-70 t^2\right) w^2\right)
\nonumber\\
&-&15 \left(4 t^2-1\right) w \left(2 \sqrt{2} u-7 w\right)\Big)+306 L_\pi^2 w^2 (L_\pi r Y_2^{\pi\pi}-4 \bar{Y}_1^{\pi\pi})\bigg]
\nonumber\\
&-&\frac{j_2(k r t)}{672 \pi  k^4 r^3 t^2} \bigg[180 L_\pi^3 r \left(2 k^2 r^2 t^2-21\right) \left(4 u^2 (Y_0^{\pi\pi}+5 Y_2^{\pi\pi})-2 \sqrt{2} u w (Y_0^{\pi\pi}+8 Y_2^{\pi\pi})+w^2 (5 Y_0^{\pi\pi}-8 )\right)
\nonumber\\
&+&180 L_\pi^4 r^2 Y_0^{\pi\pi} \left(4 u^2-2 \sqrt{2} u w+5 w^2\right) \left(2 k^2 r^2 t^2-21\right)
\nonumber\\
&+&L_\pi^2 \bar{Y}_1^{\pi\pi} \Big(-k^4 r^4 t^2 \left(336 \left(24 t^2-1\right) u^2
-4 \sqrt{2} \left(956 t^2-29\right) u w+\left(2884 t^2+2531\right) w^2\right)
\nonumber\\
&+&15 k^2 r^2 w \left(14 \sqrt{2} \left(1-4 t^2\right) u+\left(4924 t^2+1001\right) w\right)-136080 w^2\Big)
\nonumber\\
&+&144 k^2 L_\pi r t^2 w Y_0^{\pi\pi} \left(k^2 r^2 t^2 \left(14 \sqrt{2} u+141 w\right)-1680 w\right)+12 k^4 r^2 \left(4 t^2-1\right) t^2 w^2 Y_0^{\pi\pi} \left(74 k^2 r^2 t^2-1407\right)\bigg]
\nonumber\\
&-&\frac{L_\pi j_3(k r t)}{20 \pi  k^3 r^2 t} \bigg[2 k^2 r \Big(80 L_\pi^2 r^2 t^2 u^2 Y_2^{\pi\pi}-20 \sqrt{2} L_\pi r t^2 u w (2 L_\pi r Y_2^{\pi\pi}+3 \bar{Y}_1^{\pi\pi})
\nonumber\\
&+&w^2 \left(L_\pi r t^2 (169 L_\pi r Y_2^{\pi\pi}-606 \bar{Y}_1^{\pi\pi})+1215 \left(4 t^2-1\right) Y_0^{\pi\pi}\right)\Big)
\nonumber\\
&+&6 k^4 r^3 t^2 w Y_0^{\pi\pi} \left(\sqrt{2} \left(4 t^2-1\right) u+18 \left(1-5 t^2\right) w\right)-5355 L_\pi w^2 (L_\pi r Y_2^{\pi\pi}-4 \bar{Y}_1^{\pi\pi})\bigg]
\nonumber\\
&+&\frac{j_4(k r t)}{2464 \pi  k^2 r t^2} \bigg[L_\pi^2 w \bar{Y}_1^{\pi\pi} \left(k^2 r^2 t^2 \left(242 \sqrt{2} \left(4 t^2-1\right) u+\left(8347-46060 t^2\right) w\right)+99 \left(10804 t^2-2625\right) w\right)
\nonumber\\
&+&792 k^2 L_\pi r \left(4 t^2+7\right) t^2 w^2 Y_0^{\pi\pi}+132 k^4 r^2 \left(4 t^2-1\right) t^4 w^2 Y_0^{\pi\pi}+836 L_\pi^4 r^2 t^2 Y_0^{\pi\pi} \left(4 u^2-2 \sqrt{2} u w+5 w^2\right)
\nonumber\\
&+&836 L_\pi^3 r t^2 \left(4 u^2 (Y_0^{\pi\pi}+5 Y_2^{\pi\pi})-2 \sqrt{2} u w (Y_0^{\pi\pi}+8 Y_2^{\pi\pi})+w^2 (5 Y_0^{\pi\pi}-8 Y_2^{\pi\pi})\right)\bigg]
\nonumber\\
&+&\frac{3  L_\pi w^2 j_5(k r t)}{8 \pi  k t} \bigg[2 k^2 r \left(4 t^2-1\right) t^2 Y_0^{\pi\pi}+L_\pi \left(2 L_\pi r t^2 Y_2^{\pi\pi}+\left(4 t^2-3\right) \bar{Y}_1^{\pi\pi}\right)
\nonumber\\
&-&\frac{171 L_\pi^2 r \left(4 t^2-1\right) w^2 \bar{Y}_1^{\pi\pi}j_6(k r t)}{352 \pi }\bigg]
\Bigg\}
\eea
where $L_\pi=[m_\pi^2+(\frac{1}{4}-t^2)\vec{k}^2]^{1/2}$, and $Y_0^{\pi\pi}$, $\bar{Y}_1^{\pi\pi}$ and $Y_2^{\pi\pi}$ are defined as
\bea
Y_0^{\pi\pi}&=&e^{-L_\pi r}/(L_\pi r)\nonumber\\
\bar{Y}_1^{\pi\pi}&=&(1+1/(L_\pi r))Y_0^{\pi\pi}\nonumber\\
Y_2^{\pi\pi}&=&(3/(L_\pi r)^2+3/(L_\pi r)+1)Y_0^{\pi\pi}
\eea
\end{subequations}
\end{widetext}

\bibliography{refdeuteron}

\begin{thebibliography}{20}%
\makeatletter
\providecommand \@ifxundefined [1]{%
 \@ifx{#1\undefined}
}%
\providecommand \@ifnum [1]{%
 \ifnum #1\expandafter \@firstoftwo
 \else \expandafter \@secondoftwo
 \fi
}%
\providecommand \@ifx [1]{%
 \ifx #1\expandafter \@firstoftwo
 \else \expandafter \@secondoftwo
 \fi
}%
\providecommand \natexlab [1]{#1}%
\providecommand \enquote  [1]{``#1''}%
\providecommand \bibnamefont  [1]{#1}%
\providecommand \bibfnamefont [1]{#1}%
\providecommand \citenamefont [1]{#1}%
\providecommand \href@noop [0]{\@secondoftwo}%
\providecommand \href [0]{\begingroup \@sanitize@url \@href}%
\providecommand \@href[1]{\@@startlink{#1}\@@href}%
\providecommand \@@href[1]{\endgroup#1\@@endlink}%
\providecommand \@sanitize@url [0]{\catcode `\\12\catcode `\$12\catcode
  `\&12\catcode `\#12\catcode `\^12\catcode `\_12\catcode `\%12\relax}%
\providecommand \@@startlink[1]{}%
\providecommand \@@endlink[0]{}%
\providecommand \url  [0]{\begingroup\@sanitize@url \@url }%
\providecommand \@url [1]{\endgroup\@href {#1}{\urlprefix }}%
\providecommand \urlprefix  [0]{URL }%
\providecommand \Eprint [0]{\href }%
\providecommand \doibase [0]{http://dx.doi.org/}%
\providecommand \selectlanguage [0]{\@gobble}%
\providecommand \bibinfo  [0]{\@secondoftwo}%
\providecommand \bibfield  [0]{\@secondoftwo}%
\providecommand \translation [1]{[#1]}%
\providecommand \BibitemOpen [0]{}%
\providecommand \bibitemStop [0]{}%
\providecommand \bibitemNoStop [0]{.\EOS\space}%
\providecommand \EOS [0]{\spacefactor3000\relax}%
\providecommand \BibitemShut  [1]{\csname bibitem#1\endcsname}%
\let\auto@bib@innerbib\@empty
\bibitem [{\citenamefont {Freese}\ and\ \citenamefont
  {Cosyn}(2022)}]{Freese:2022yur}%
  \BibitemOpen
  \bibfield  {author} {\bibinfo {author} {\bibfnamefont {Adam}\ \bibnamefont
  {Freese}}\ and\ \bibinfo {author} {\bibfnamefont {Wim}\ \bibnamefont
  {Cosyn}},\ }\bibfield  {title} {\enquote {\bibinfo {title} {{Spatial
  densities of momentum and forces in spin-one hadrons}},}\ }\href {\doibase
  10.1103/PhysRevD.106.114013} {\bibfield  {journal} {\bibinfo  {journal}
  {Phys. Rev. D}\ }\textbf {\bibinfo {volume} {106}},\ \bibinfo {pages}
  {114013} (\bibinfo {year} {2022})},\ \Eprint
  {http://arxiv.org/abs/2207.10787} {arXiv:2207.10787 [hep-ph]} \BibitemShut
  {NoStop}%
\bibitem [{\citenamefont {Garcia Martin-Caro}\ \emph
  {et~al.}(2023)\citenamefont {Garcia Martin-Caro}, \citenamefont {Huidobro},\
  and\ \citenamefont {Hatta}}]{GarciaMartin-Caro:2023klo}%
  \BibitemOpen
  \bibfield  {author} {\bibinfo {author} {\bibfnamefont {Alberto}\ \bibnamefont
  {Garcia Martin-Caro}}, \bibinfo {author} {\bibfnamefont {Miguel}\
  \bibnamefont {Huidobro}}, \ and\ \bibinfo {author} {\bibfnamefont
  {Yoshitaka}\ \bibnamefont {Hatta}},\ }\bibfield  {title} {\enquote {\bibinfo
  {title} {{Gravitational form factors of nuclei in the Skyrme model}},}\
  }\href {\doibase 10.1103/PhysRevD.108.034014} {\bibfield  {journal} {\bibinfo
   {journal} {Phys. Rev. D}\ }\textbf {\bibinfo {volume} {108}},\ \bibinfo
  {pages} {034014} (\bibinfo {year} {2023})},\ \Eprint
  {http://arxiv.org/abs/2304.05994} {arXiv:2304.05994 [nucl-th]} \BibitemShut
  {NoStop}%
\bibitem [{\citenamefont {He}\ and\ \citenamefont {Zahed}(2023)}]{He:2023ogg}%
  \BibitemOpen
  \bibfield  {author} {\bibinfo {author} {\bibfnamefont {Fangcheng}\
  \bibnamefont {He}}\ and\ \bibinfo {author} {\bibfnamefont {Ismail}\
  \bibnamefont {Zahed}},\ }\bibfield  {title} {\enquote {\bibinfo {title}
  {{Gravitational form factors of light nuclei: Impulse approximation}},}\
  }\href@noop {} {\  (\bibinfo {year} {2023})},\ \Eprint
  {http://arxiv.org/abs/2310.12315} {arXiv:2310.12315 [nucl-th]} \BibitemShut
  {NoStop}%
\bibitem [{\citenamefont {Hockert}\ \emph {et~al.}(1973)\citenamefont
  {Hockert}, \citenamefont {Riska}, \citenamefont {Gari},\ and\ \citenamefont
  {Huffman}}]{Hockert:1973fot}%
  \BibitemOpen
  \bibfield  {author} {\bibinfo {author} {\bibfnamefont {J.}~\bibnamefont
  {Hockert}}, \bibinfo {author} {\bibfnamefont {D.~O.}\ \bibnamefont {Riska}},
  \bibinfo {author} {\bibfnamefont {M.}~\bibnamefont {Gari}}, \ and\ \bibinfo
  {author} {\bibfnamefont {A.}~\bibnamefont {Huffman}},\ }\bibfield  {title}
  {\enquote {\bibinfo {title} {{Meson exchange currents in deuteron
  electrodisintegration}},}\ }\href {\doibase 10.1016/0375-9474(73)90621-0}
  {\bibfield  {journal} {\bibinfo  {journal} {Nucl. Phys. A}\ }\textbf
  {\bibinfo {volume} {217}},\ \bibinfo {pages} {14--28} (\bibinfo {year}
  {1973})}\BibitemShut {NoStop}%
\bibitem [{\citenamefont {Chemtob}\ \emph {et~al.}(1974)\citenamefont
  {Chemtob}, \citenamefont {Moniz},\ and\ \citenamefont
  {Rho}}]{Chemtob:1974nf}%
  \BibitemOpen
  \bibfield  {author} {\bibinfo {author} {\bibfnamefont {M.}~\bibnamefont
  {Chemtob}}, \bibinfo {author} {\bibfnamefont {E.~J.}\ \bibnamefont {Moniz}},
  \ and\ \bibinfo {author} {\bibfnamefont {Mannque}\ \bibnamefont {Rho}},\
  }\bibfield  {title} {\enquote {\bibinfo {title} {{Deuteron Electromagnetic
  Structure at Large Momentum Transfer}},}\ }\href {\doibase
  10.1103/PhysRevC.10.344} {\bibfield  {journal} {\bibinfo  {journal} {Phys.
  Rev. C}\ }\textbf {\bibinfo {volume} {10}},\ \bibinfo {pages} {344--352}
  (\bibinfo {year} {1974})}\BibitemShut {NoStop}%
\bibitem [{\citenamefont {Kloet}\ and\ \citenamefont
  {Tjon}(1974)}]{Kloet:1973mj}%
  \BibitemOpen
  \bibfield  {author} {\bibinfo {author} {\bibfnamefont {W.~M.}\ \bibnamefont
  {Kloet}}\ and\ \bibinfo {author} {\bibfnamefont {J.~A.}\ \bibnamefont
  {Tjon}},\ }\bibfield  {title} {\enquote {\bibinfo {title} {{MESON EXCHANGE
  EFFECTS ON THE CHARGE FORM-FACTORS OF THE TRINUCLEON SYSTEM}},}\ }\href
  {\doibase 10.1016/0370-2693(74)90623-6} {\bibfield  {journal} {\bibinfo
  {journal} {Phys. Lett. B}\ }\textbf {\bibinfo {volume} {49}},\ \bibinfo
  {pages} {419--422} (\bibinfo {year} {1974})}\BibitemShut {NoStop}%
\bibitem [{\citenamefont {Jackson}\ \emph {et~al.}(1975)\citenamefont
  {Jackson}, \citenamefont {Lande},\ and\ \citenamefont
  {Riska}}]{Jackson:1975fys}%
  \BibitemOpen
  \bibfield  {author} {\bibinfo {author} {\bibfnamefont {A.~D.}\ \bibnamefont
  {Jackson}}, \bibinfo {author} {\bibfnamefont {A.}~\bibnamefont {Lande}}, \
  and\ \bibinfo {author} {\bibfnamefont {D.~O.}\ \bibnamefont {Riska}},\
  }\bibfield  {title} {\enquote {\bibinfo {title} {{Pion exchange currents in
  elastic electron deuteron scattering}},}\ }\href {\doibase
  10.1016/0370-2693(75)90177-X} {\bibfield  {journal} {\bibinfo  {journal}
  {Phys. Lett. B}\ }\textbf {\bibinfo {volume} {55}},\ \bibinfo {pages}
  {23--27} (\bibinfo {year} {1975})}\BibitemShut {NoStop}%
\bibitem [{\citenamefont {Alharazin}\ \emph {et~al.}(2020)\citenamefont
  {Alharazin}, \citenamefont {Djukanovic}, \citenamefont {Gegelia},\ and\
  \citenamefont {Polyakov}}]{Alharazin:2020yjv}%
  \BibitemOpen
  \bibfield  {author} {\bibinfo {author} {\bibfnamefont {H.}~\bibnamefont
  {Alharazin}}, \bibinfo {author} {\bibfnamefont {D.}~\bibnamefont
  {Djukanovic}}, \bibinfo {author} {\bibfnamefont {J.}~\bibnamefont {Gegelia}},
  \ and\ \bibinfo {author} {\bibfnamefont {M.~V.}\ \bibnamefont {Polyakov}},\
  }\bibfield  {title} {\enquote {\bibinfo {title} {{Chiral theory of nucleons
  and pions in the presence of an external gravitational field}},}\ }\href
  {\doibase 10.1103/PhysRevD.102.076023} {\bibfield  {journal} {\bibinfo
  {journal} {Phys. Rev. D}\ }\textbf {\bibinfo {volume} {102}},\ \bibinfo
  {pages} {076023} (\bibinfo {year} {2020})},\ \Eprint
  {http://arxiv.org/abs/2006.05890} {arXiv:2006.05890 [hep-ph]} \BibitemShut
  {NoStop}%
\bibitem [{\citenamefont {Duran}\ \emph {et~al.}(2023)\citenamefont {Duran}
  \emph {et~al.}}]{Duran:2022xag}%
  \BibitemOpen
  \bibfield  {author} {\bibinfo {author} {\bibfnamefont {B.}~\bibnamefont
  {Duran}} \emph {et~al.},\ }\bibfield  {title} {\enquote {\bibinfo {title}
  {{Determining the gluonic gravitational form factors of the proton}},}\
  }\href {\doibase 10.1038/s41586-023-05730-4} {\bibfield  {journal} {\bibinfo
  {journal} {Nature}\ }\textbf {\bibinfo {volume} {615}},\ \bibinfo {pages}
  {813--816} (\bibinfo {year} {2023})},\ \Eprint
  {http://arxiv.org/abs/2207.05212} {arXiv:2207.05212 [nucl-ex]} \BibitemShut
  {NoStop}%
\bibitem [{\citenamefont {Hackett}\ \emph
  {et~al.}(2023{\natexlab{a}})\citenamefont {Hackett}, \citenamefont {Pefkou},\
  and\ \citenamefont {Shanahan}}]{Hackett:2023rif}%
  \BibitemOpen
  \bibfield  {author} {\bibinfo {author} {\bibfnamefont {Daniel~C.}\
  \bibnamefont {Hackett}}, \bibinfo {author} {\bibfnamefont {Dimitra~A.}\
  \bibnamefont {Pefkou}}, \ and\ \bibinfo {author} {\bibfnamefont {Phiala~E.}\
  \bibnamefont {Shanahan}},\ }\bibfield  {title} {\enquote {\bibinfo {title}
  {{Gravitational form factors of the proton from lattice QCD}},}\ }\href@noop
  {} {\  (\bibinfo {year} {2023}{\natexlab{a}})},\ \Eprint
  {http://arxiv.org/abs/2310.08484} {arXiv:2310.08484 [hep-lat]} \BibitemShut
  {NoStop}%
\bibitem [{\citenamefont {Guo}\ \emph {et~al.}(2023)\citenamefont {Guo},
  \citenamefont {Ji}, \citenamefont {Liu},\ and\ \citenamefont
  {Yang}}]{Guo:2023pqw}%
  \BibitemOpen
  \bibfield  {author} {\bibinfo {author} {\bibfnamefont {Yuxun}\ \bibnamefont
  {Guo}}, \bibinfo {author} {\bibfnamefont {Xiangdong}\ \bibnamefont {Ji}},
  \bibinfo {author} {\bibfnamefont {Yizhuang}\ \bibnamefont {Liu}}, \ and\
  \bibinfo {author} {\bibfnamefont {Jinghong}\ \bibnamefont {Yang}},\
  }\bibfield  {title} {\enquote {\bibinfo {title} {{Updated analysis of
  near-threshold heavy quarkonium production for probe of
  proton\textquoteright{}s gluonic gravitational form factors}},}\ }\href
  {\doibase 10.1103/PhysRevD.108.034003} {\bibfield  {journal} {\bibinfo
  {journal} {Phys. Rev. D}\ }\textbf {\bibinfo {volume} {108}},\ \bibinfo
  {pages} {034003} (\bibinfo {year} {2023})},\ \Eprint
  {http://arxiv.org/abs/2305.06992} {arXiv:2305.06992 [hep-ph]} \BibitemShut
  {NoStop}%
\bibitem [{\citenamefont {Mamo}\ and\ \citenamefont
  {Zahed}(2020)}]{Mamo:2019mka}%
  \BibitemOpen
  \bibfield  {author} {\bibinfo {author} {\bibfnamefont {Kiminad~A.}\
  \bibnamefont {Mamo}}\ and\ \bibinfo {author} {\bibfnamefont {Ismail}\
  \bibnamefont {Zahed}},\ }\bibfield  {title} {\enquote {\bibinfo {title}
  {{Diffractive photoproduction of $J/\psi$ and $\Upsilon$ using holographic
  QCD: gravitational form factors and GPD of gluons in the proton}},}\ }\href
  {\doibase 10.1103/PhysRevD.101.086003} {\bibfield  {journal} {\bibinfo
  {journal} {Phys. Rev. D}\ }\textbf {\bibinfo {volume} {101}},\ \bibinfo
  {pages} {086003} (\bibinfo {year} {2020})},\ \Eprint
  {http://arxiv.org/abs/1910.04707} {arXiv:1910.04707 [hep-ph]} \BibitemShut
  {NoStop}%
\bibitem [{\citenamefont {Mamo}\ and\ \citenamefont
  {Zahed}(2022)}]{Mamo:2022eui}%
  \BibitemOpen
  \bibfield  {author} {\bibinfo {author} {\bibfnamefont {Kiminad~A.}\
  \bibnamefont {Mamo}}\ and\ \bibinfo {author} {\bibfnamefont {Ismail}\
  \bibnamefont {Zahed}},\ }\bibfield  {title} {\enquote {\bibinfo {title}
  {{J/\ensuremath{\psi} near threshold in holographic QCD: A and D
  gravitational form factors}},}\ }\href {\doibase 10.1103/PhysRevD.106.086004}
  {\bibfield  {journal} {\bibinfo  {journal} {Phys. Rev. D}\ }\textbf {\bibinfo
  {volume} {106}},\ \bibinfo {pages} {086004} (\bibinfo {year} {2022})},\
  \Eprint {http://arxiv.org/abs/2204.08857} {arXiv:2204.08857 [hep-ph]}
  \BibitemShut {NoStop}%
\bibitem [{\citenamefont {Polyakov}\ and\ \citenamefont
  {Sun}(2019)}]{Polyakov:2019lbq}%
  \BibitemOpen
  \bibfield  {author} {\bibinfo {author} {\bibfnamefont {Maxim~V.}\
  \bibnamefont {Polyakov}}\ and\ \bibinfo {author} {\bibfnamefont {Bao-Dong}\
  \bibnamefont {Sun}},\ }\bibfield  {title} {\enquote {\bibinfo {title}
  {{Gravitational form factors of a spin one particle}},}\ }\href {\doibase
  10.1103/PhysRevD.100.036003} {\bibfield  {journal} {\bibinfo  {journal}
  {Phys. Rev. D}\ }\textbf {\bibinfo {volume} {100}},\ \bibinfo {pages}
  {036003} (\bibinfo {year} {2019})},\ \Eprint
  {http://arxiv.org/abs/1903.02738} {arXiv:1903.02738 [hep-ph]} \BibitemShut
  {NoStop}%
\bibitem [{\citenamefont {Gari}\ and\ \citenamefont
  {Hyuga}(1976)}]{Gari:1976kj}%
  \BibitemOpen
  \bibfield  {author} {\bibinfo {author} {\bibfnamefont {M.}~\bibnamefont
  {Gari}}\ and\ \bibinfo {author} {\bibfnamefont {H.}~\bibnamefont {Hyuga}},\
  }\bibfield  {title} {\enquote {\bibinfo {title} {{Mesonic Degrees of Freedom
  in Nuclei and the Definition of Meson Exchange Currents}},}\ }\href {\doibase
  10.1007/BF01415604} {\bibfield  {journal} {\bibinfo  {journal} {Z. Phys. A}\
  }\textbf {\bibinfo {volume} {277}},\ \bibinfo {pages} {291--297} (\bibinfo
  {year} {1976})}\BibitemShut {NoStop}%
\bibitem [{\citenamefont {Hackett}\ \emph
  {et~al.}(2023{\natexlab{b}})\citenamefont {Hackett}, \citenamefont {Oare},
  \citenamefont {Pefkou},\ and\ \citenamefont {Shanahan}}]{Hackett:2023nkr}%
  \BibitemOpen
  \bibfield  {author} {\bibinfo {author} {\bibfnamefont {Daniel~C.}\
  \bibnamefont {Hackett}}, \bibinfo {author} {\bibfnamefont {Patrick~R.}\
  \bibnamefont {Oare}}, \bibinfo {author} {\bibfnamefont {Dimitra~A.}\
  \bibnamefont {Pefkou}}, \ and\ \bibinfo {author} {\bibfnamefont {Phiala~E.}\
  \bibnamefont {Shanahan}},\ }\bibfield  {title} {\enquote {\bibinfo {title}
  {{Gravitational form factors of the pion from lattice QCD}},}\ }\href
  {\doibase 10.1103/PhysRevD.108.114504} {\bibfield  {journal} {\bibinfo
  {journal} {Phys. Rev. D}\ }\textbf {\bibinfo {volume} {108}},\ \bibinfo
  {pages} {114504} (\bibinfo {year} {2023}{\natexlab{b}})},\ \Eprint
  {http://arxiv.org/abs/2307.11707} {arXiv:2307.11707 [hep-lat]} \BibitemShut
  {NoStop}%
\bibitem [{\citenamefont {Kaptari}\ \emph {et~al.}(1990)\citenamefont
  {Kaptari}, \citenamefont {Titov}, \citenamefont {Bratkovskaya},\ and\
  \citenamefont {Umnikov}}]{Kaptari:1989un}%
  \BibitemOpen
  \bibfield  {author} {\bibinfo {author} {\bibfnamefont {L.~P.}\ \bibnamefont
  {Kaptari}}, \bibinfo {author} {\bibfnamefont {A.~I.}\ \bibnamefont {Titov}},
  \bibinfo {author} {\bibfnamefont {E.~L.}\ \bibnamefont {Bratkovskaya}}, \
  and\ \bibinfo {author} {\bibfnamefont {A.~Yu.}\ \bibnamefont {Umnikov}},\
  }\bibfield  {title} {\enquote {\bibinfo {title} {{Meson Exchange Corrections
  in Deep Inelastic Scattering on Deuteron}},}\ }\href {\doibase
  10.1016/0375-9474(90)90230-J} {\bibfield  {journal} {\bibinfo  {journal}
  {Nucl. Phys. A}\ }\textbf {\bibinfo {volume} {512}},\ \bibinfo {pages}
  {684--698} (\bibinfo {year} {1990})}\BibitemShut {NoStop}%
\bibitem [{\citenamefont {Ohta}\ and\ \citenamefont
  {Wakamatsu}(1976)}]{Ohta:1975qv}%
  \BibitemOpen
  \bibfield  {author} {\bibinfo {author} {\bibfnamefont {Kohichi}\ \bibnamefont
  {Ohta}}\ and\ \bibinfo {author} {\bibfnamefont {Masashi}\ \bibnamefont
  {Wakamatsu}},\ }\bibfield  {title} {\enquote {\bibinfo {title} {{Nucleon
  Recoil Current in the Nuclear Weak and Electromagnetic Interactions}},}\
  }\href {\doibase 10.1143/PTP.55.131} {\bibfield  {journal} {\bibinfo
  {journal} {Prog. Theor. Phys.}\ }\textbf {\bibinfo {volume} {55}},\ \bibinfo
  {pages} {131} (\bibinfo {year} {1976})}\BibitemShut {NoStop}%
\bibitem [{\citenamefont {Siegert}(1937)}]{Siegert:1937yt}%
  \BibitemOpen
  \bibfield  {author} {\bibinfo {author} {\bibfnamefont {A.~J.~F.}\
  \bibnamefont {Siegert}},\ }\bibfield  {title} {\enquote {\bibinfo {title}
  {{Note on the interaction between nuclei and electromagnetic radiation}},}\
  }\href {\doibase 10.1103/PhysRev.52.787} {\bibfield  {journal} {\bibinfo
  {journal} {Phys. Rev.}\ }\textbf {\bibinfo {volume} {52}},\ \bibinfo {pages}
  {787--789} (\bibinfo {year} {1937})}\BibitemShut {NoStop}%
\bibitem [{\citenamefont {Austern}\ and\ \citenamefont
  {Sachs}(1951)}]{Austern:1951zz}%
  \BibitemOpen
  \bibfield  {author} {\bibinfo {author} {\bibfnamefont {N.}~\bibnamefont
  {Austern}}\ and\ \bibinfo {author} {\bibfnamefont {R.~G.}\ \bibnamefont
  {Sachs}},\ }\bibfield  {title} {\enquote {\bibinfo {title} {{Interaction
  Effects on Radiative Transitions in Nuclei}},}\ }\href {\doibase
  10.1103/PhysRev.81.710} {\bibfield  {journal} {\bibinfo  {journal} {Phys.
  Rev.}\ }\textbf {\bibinfo {volume} {81}},\ \bibinfo {pages} {710--716}
  (\bibinfo {year} {1951})}\BibitemShut {NoStop}%
\end{thebibliography}%
\end{document}